\documentclass[11pt,a4paper]{article}

\usepackage[truedimen,margin=30mm]{geometry} 

\usepackage{mathrsfs}
\usepackage{amssymb}
\usepackage{amsmath}
\usepackage{ascmac}
\usepackage{amsthm}
\usepackage{amsmath}
\usepackage[dvipdfmx]{graphicx}
\usepackage[]{color}
\usepackage{booktabs}
\usepackage{setspace}
\usepackage{natbib}
\usepackage{float}
\usepackage{mathtools}
\usepackage{url}

\usepackage{titlesec}
\titleformat*{\section}{\large\bfseries}
\titleformat*{\subsection}{\it}

\def\zero{{\text{\boldmath$0$}}}

\def\a{{\text{\boldmath $a$}}}
\def\e{{\text{\boldmath $e$}}}
\def\d{{\text{\boldmath $d$}}}
\def\f{{\text{\boldmath $f$}}}
\def\g{{\text{\boldmath $g$}}}
\def\m{{\text{\boldmath $m$}}}

\def\v{{\text{\boldmath $v$}}}

\def\x{{\text{\boldmath $x$}}}

\def\z{{\text{\boldmath $z$}}}

\def\A{{\text{\boldmath $A$}}}

\def\C{{\text{\boldmath $C$}}}
\def\F{{\text{\boldmath $F$}}}

\def\I{{\text{\boldmath $I$}}}
\def\F{{\text{\boldmath $F$}}}

\def\Q{{\text{\boldmath $Q$}}}

\def\R{{\text{\boldmath $R$}}}
\def\S{{\text{\boldmath $S$}}}
\def\u{{\text{\boldmath $u$}}}
\def\W{{\text{\boldmath $W$}}}

\def\Z{{\text{\boldmath $Z$}}}

\def\bbe{{\text{\boldmath $\beta$}}}

\def\bpi{{\text{\boldmath $\pi$}}}
\def\bthe{{\text{\boldmath $\theta$}}}
\def\bPhi{{\text{\boldmath $\Phi$}}}
\def\bOmega{{\text{\boldmath $\Omega$}}}
\def\bSig{{\text{\boldmath $\Sigma$}}}

\def\diag{\text{diag}}

\title{\Large{\textbf{Predicting COVID-19 hospitalisation using a mixture of Bayesian predictive syntheses}
}}

\date{}
\author{
}

\begin{document}

\maketitle
\doublespacing

\vspace{-1.5cm}
\begin{center}
Genya Kobayashi$^{1}$\footnote{Author of correspondence: \texttt{gkobayashi@meiji.ac.jp}}, Shonosuke Sugasawa$^{2}$, Yuki Kawakubo$^{3}$, \\ 
{Dongu Han}$^4$ and Taeryon Choi$^4$

\end{center}

\noindent
$^1$School of Commerce, Meiji University, Japan\\
$^2$Faculty of Economics, Keio University, Japan\\
$^3$Graduate School of Social Sciences, Chiba University, Japan\\
$^4$Department of Statistics, Korea University, Korea

\medskip
\begin{center}
{\bf  Abstract}
\end{center}
This paper proposes a novel methodology called the mixture of Bayesian predictive syntheses (MBPS) for multiple time series count data for the challenging task of predicting the numbers of COVID-19 inpatients and isolated cases in Japan and Korea at the subnational-level. 
MBPS combines a set of predictive models and partitions the multiple time series into clusters based on their contribution to predicting the outcome.
In this way, MBPS leverages the shared information within each cluster and is suitable for predicting COVID-19 inpatients since the data exhibit similar dynamics over multiple areas. 
Also, MBPS avoids using a multivariate count model, which is generally cumbersome to develop and implement. 
Our Japanese and Korean data analyses demonstrate that the proposed MBPS methodology has improved predictive accuracy and uncertainty quantification. 
\vspace{-0cm}


\section{Introduction}
The outbreak of COVID-19 caused tremendous social and economic disruption all over the world. 
During the pandemic that the modern world had not experienced, there were explosions in the various academic fields to investigate and predict the various aspects of the pandemic and its influence.

In Japan, COVID-19 was classified as Category II Infectious Diseases under the Infectious Diseases Control Law until 7 May 2023. 
During the pandemic, infected individuals were isolated, and patients with severe conditions received treatment in designated beds. 
The number of designated beds is limited and varied across the prefectures of Japan. 
In the fall and winter of 2021, when the Omicron variant  (BA-5 virus) spread dramatically, there was a deep concern about the shortage of designated beds for COVID-19 patients. 
In Korea, COVID-19 was classified as a first-grade infectious disease by the Korea Centers for Disease Control and Prevention (KCDC) until 24 April 2022. 
As in Japan, infected individuals underwent isolation or hospitalisation depending on the severity of the condition. 
The emergence of the Omicron variant also led to a significant increase in isolated individuals. 
These backgrounds highlight the need for quick responses by national and local governments to rapid changes in healthcare demand and for efficient resource allocation, which are based on accurate patient prediction, desirably, at the subnational level. 

Hence, the aim of this paper is to provide a novel statistical method that performs reasonably well in predicting the numbers of inpatients and isolated cases in Japan and Korea at the subnational level. 
From the early stage of the pandemic, many studies tried to predict its course or extract the trend using existing and new methods \citep[see][]{Rahimi}. 
However, due to the very complex and unexpected dynamics of the pandemic caused by the rise of new variants, government interventions, and various formal and informal prevention measures, it is not possible for a single model to consistently outperform others. 
Therefore, instead of implementing a single overly complex model to predict the course of the pandemic, combining predictions from a set of relatively simple predictive models would be more fruitful, as demonstrated by \cite{Paireau} and \cite{Chowell}, which performed an ensemble prediction for several indicators of the country-level healthcare demand in France and the United States, respectively.

Therefore, we develop a novel statistical method that combines multiple predictive models, specifically focusing on the emerging approach called Bayesian predictive synthesis (BPS) proposed by \cite{MW19}. 
BPS is a coherent Bayesian framework for synthesising multiple predictive models, called agent models, based on the agent opinion analysis of, for example, \cite{Gene} and \cite{WC92}.  
The recent studies on BPS include \cite{McAlinn20} for multivariate time series, \cite{mcalinn2021mixed} for mixed-frequency time series, \cite{cabel2022spatially} for spatial data and \cite{sugasawa23} for the meta-inference of heterogeneous treatment effects. 
Notably, \cite{TM23} showed that BPS is exact minimax for time series analysis.  
See \cite{cabel2022spatially} also for minimaxity of BPS for spatial data. 
\cite{tallman2022bayesian} considered BPS in the general Bayes setting. 
\cite{JW23} proposed several ways to construct synthesis weights that depend on the outcomes. 
\cite{Chernis} considered the situation where an analyst has many agent models to be combined and proposed the shrinkage approach to suppress the contribution of redundant models and the factor approach to reduce the number of latent variables. 
See also \cite{Aastveit19} for a review of other forecast combination methods and references therein.

Since the numbers of inpatients and isolated cases in Japan and Korea constitute time series count data over multiple areas of the countries, we propose a novel BPS methodology for multiple time series count data called the mixture of Bayesian predictive syntheses (MBPS).  
In our analysis, we have the count time series for 47 prefectures of Japan and 17 Metropolitan Autonomous Governments (MAGs) of Korea. 
Instead of treating multiple count time series using multivariate models, the proposed MBPS partitions them into clusters based on the finite mixture of BPS  since the specification and implementation of multivariate count time series models are generally cumbersome \citep{Davis21, Fokianos21, West20}.  
Introducing a clustering structure is motivated by the observation that a group of time series in our data exhibits similar dynamics. 
In MBPS, the agent models are assigned the same synthesis weights for the time series in the cluster. 
In this way, MBPS can leverage the shared information across time series based on the similarity in the contribution of the agent models to predicting the outcomes. 
Our MBPS is similar to the time series clustering methods based on the parametric models reviewed in \cite{FS11} and the nonparametric Bayesian approach of  \cite{NC14} and \cite{LIN19}, in the sense that they have cluster-specific parameters and latent variables. 
Note that the existing BPS approaches do not consider such a clustering structure. 
The clustering structure of MBPS can reduce the number of parameters and model complexity compared to a fully multivariate count model. 
Each agent model fed into MBPS can be univariate, and the individual univariate predictions are combined through the mixture of synthesised predictive distributions. 
It is appealing because there is an abundance of univariate count models \citep{Davis16, Davis21, Fokianos21}. 
Thus, one can easily choose and fit a set of univariate models used in MBPS.

The remainder of this paper is organised as follows. 
Section~\ref{sec:data} introduces the data on the number of COVID-19 inpatients in the prefectures of Japan and number of isolated cases in the MAGs of Korea. 
Section~\ref{sec:method} first briefly describes BPS. 
Then, we propose the novel BPS for time series count data and introduce  MBPS for multiple time series count data and its variants. 
Using MBPS, Section~\ref{sec:real} analyses the COVID-19-related data on the numbers of inpatients in Japan and isolated cases in Korea, where the proposed method demonstrates improved predictive accuracy and uncertainty quantification. 
Finally, Section~\ref{sec:conc} provides some conclusion and discussion.

\section{Data}\label{sec:data}
\begin{figure}[H]
\centering
\includegraphics[scale=0.35]{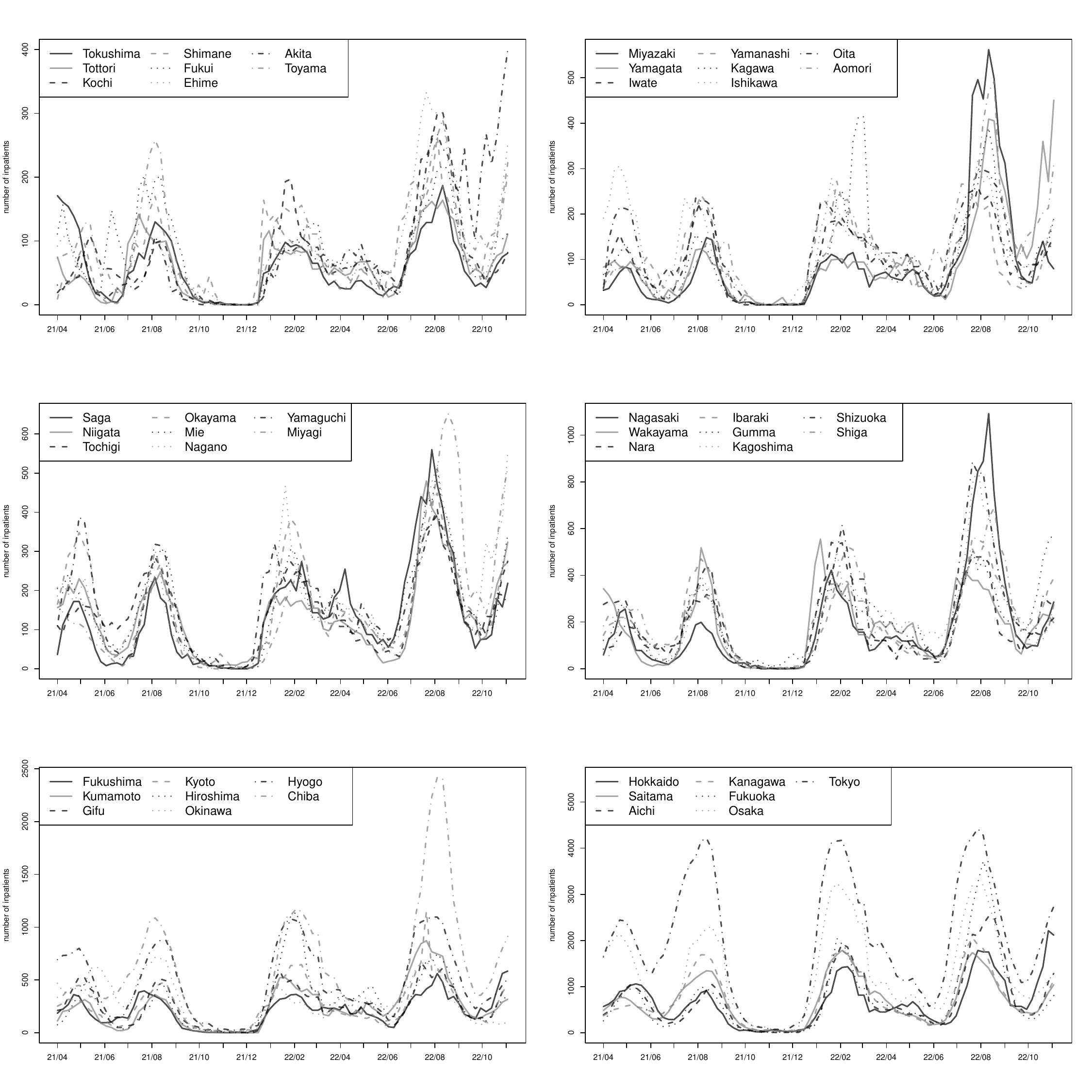}
\caption{Time series plots of the numbers of inpatients in Japan. The prefectures are grouped based on the means of the inpatients over the data period. }
\label{fig:jp}
\end{figure}

We obtained the weekly numbers of COVID-19 inpatients in the 47 prefectures of Japan from the Ministry of Health, Labour and Welfare of Japan (\url{https://www.mhlw.go.jp}). 
The data are publicly available. 
The total data period is from 7 May 2020 to 23 November 2022 (134 weeks). 
Figure~\ref{fig:jp} shows the time series plots between 21 April 2021 and 23 November 2022. 
In the figure, the prefectures are grouped arbitrarily into each panel based on the average number of inpatients over the data period for visibility. 
In the figure, the dynamics among the time series are very similar, with surges as the new variants prevailed and drops as the immunity to them was gained.

\begin{figure}[H]
\centering
\includegraphics[width=\textwidth]{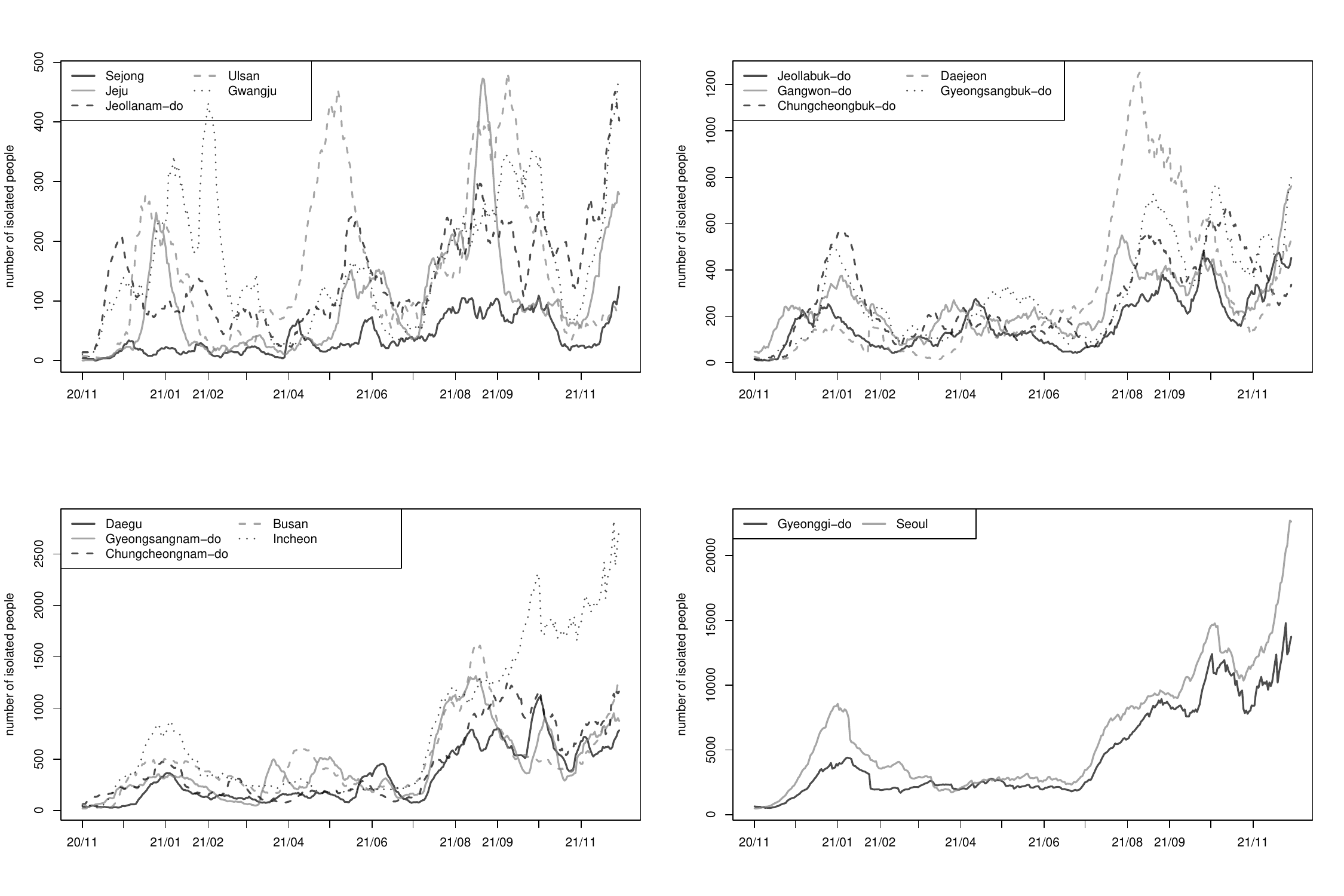}
\caption{Time series plots of the number of isolated people in Korea. The MAGs are grouped based on the averages of the isolated people over the data period. }
\label{fig:kr}
\end{figure}

In Korea, KCDC provides the open data via the public data portal (\url{https://www.data.go.kr}). 
The dataset consists of the number of daily isolated cases of COVID-19 from the 17 MAG, spanning from 1 August 2020 to 30 November 2021, totalling 487 days. 
We could not access additional data beyond this period as the Korean government ceased publishing the number of daily isolated cases starting in December 2021. 
Figure~\ref{fig:kr} presents the time series plot of the Korean data over the estimation and prediction periods where MAGs are grouped based on the averages of the isolated people over the data period. 
As in the Japanese data, the data exhibit similar dynamics among the series.

\subsection{Bayesian predictive synthesis}
Consider a univariate time series data denoted by $y_t$ for $t=1,2,\dots$. 
An analyst elicits $J$ agent models with the predictive density denoted by $h_{tj}(f_{tj})$ for $j=1,\dots,J$. 
Then to predict $y_t$ at the time $t-1$, the collection of $J$ predictive densities forms an information set denoted by $\mathcal{H}_t=(h_{t1}(y_t),\dots,h_{tJ}(y_t))$. 
The available information is $\left\{y_{1:t-1}, \mathcal{H}_{1:t}\right\}$. 
The Bayesian predictive synthesis considers the following synthesised predictive distribution \citep{MW19}:
\[
    p(y_t|\bPhi_t,y_{1:t-1},\mathcal{H}_{1:t})\equiv p(y_t|\bPhi_t,\mathcal{H}_t)=\int\alpha(y_t|\f_t,\bPhi_t)\left[\prod_{j=1}^J h_{tj}(f_{tj})\right]d\f_t, 
\]
where $\alpha(y_t|\f_t,\bPhi_t)$ denotes the synthesis function which controls how to combine $J$ predictions from the agent models,  $\bPhi_t$ is the collection of parameters of the synthesis function, $\f_t=(f_{t1},\dots,f_{tJ})'$ denotes the draw from the predictive densities that are regarded as the latent factors.  
The design of the synthesis function and prior structure of the synthesis weights are left to the analyst's discretion. 
Thus, the model can vary depending on the purpose of the analysis and the type and potential dependence structure of the data. 

The distribution of the latent factor for $j$th agent model $h_{tj}$ may be typically assumed to be the normal distribution and we set $h_{tj}(f_{tj})=\phi(f_{tj};m_{tj},s_{tj}^2)$ for $j=1,\dots,J$. 
The agent models are fitted independently before the BPS model. 
Then, the mean and variance of $h_{tj}$, $m_{tj}$ and $s_{tj}^2$, are obtained exactly or by analytic or numerical approximation for each agent. 
If $j$th agent model is Bayesian, $m_{tj}$ and $s_{tj}^2$, respectively, would correspond to the mean and variance of the posterior predictive distribution of $y_t$ given $y_{1:t}$ under this agent. 
They are obtained analytically or approximated using a Monte Carlo method. 
In the case where the agent is fitted in a frequentist framework, $m_{tj}$ and $s_{tj}^2$ can also be obtained from the predictive distribution of $y_t$ given $y_{1:t-1}$ analytically or numerically.

When $y_t$ is supported on the real line, \cite{MW19} proposed the use of the normal probability density $\alpha(y_t|\f_t,\bPhi_t)=\phi(y_t;\mu_t,\sigma^2_t)$ as the synthesis function, where $\phi(\cdot;\mu,\sigma^2)$ denotes the probability density function of $N(\mu,\sigma^2)$. 
Introducing the synthesis weights for the $J$ agent models, the mean $\mu_t$ is modelled as $\mu_t=\bthe_t'\F_t$ where  $\bthe_t=(\theta_{t0},\theta_{t1},\dots,\theta_{tJ})'$ is the $(J+1)\times 1$ vector of the synthesis weights and $\F_t=(1,\f_t')'$. 
\cite{MW19} employed the dynamic linear model (DLM) with different specifications for the agent models. 
Then $m_{tj}$ and $s_{tj}^2$ can be obtained from the standard DLM updating scheme, namely the one-step-ahead predictive distribution of  $y_t|\x_{1:t},y_{1:t-1}$ where $\x_{1:t}$ denotes the covariates information up to $t$.

Since the latent factor  $f_{tj}$ is a draw from $h_{tj}$ in BPS, the synthesised predictive model for the continuous $y_t$ can be written in the form of a normal factor model: $y_t|\bthe_t' \F_t,\sigma^2_t\sim N(\bthe_t'\F_t,\sigma^2_t)$, $f_{tj} \sim N(m_{tj},s_{tj}^2),\ j=1,\dots,J$, for $t=1,2,\dots$. 
In this case, $\bPhi_t=(\bthe_t,\sigma^2_t)$. 
Furthermore, an appropriate model for $\bthe_t$ can be specified; see Section~\ref{sec:mbps}. 
The synthesis weights $\bthe_t$ are not constrained to be in the $J$-simplex and allow negative weights. 
Thus, when $\theta_{tj}$ for some $j$ is close to zero, the corresponding agent model does not contribute to predicting $y_t$. 
The intercept $\theta_{t0}$ captures the remaining part of predicting $y_t$ and can be interpreted as the inadequacy of the predictive performance of the agent models.

When $y_t$ is the univariate count data, we propose to use the Poisson probability mass function for $\alpha(y_t|\f_t,\bPhi_t)$ with $\bPhi_t=\bthe_t$. 
Therefore, conditionally on the latent factors  $\f_t$ and synthesis weights $\bthe_t$, we have 
\begin{equation}\label{eqn:Po_uni}
y_t|\bthe_t,\f_t \sim Poi(\lambda_{t}), \quad \lambda_t=\exp(\bthe_t'\F_t).  
\end{equation}
To our best knowledge, this is the first study to treat the discrete time series within the BPS framework. 
Unlike the normal factor BPS model, the latent factors are in the exponential function. 
Since our COVID-19 hospitalisation data also include large counts, to avoid numerical instability, the predictive distribution of $\log y_t$ is used to construct $h_{tj}$. 
Then, the normal density is used for $h_{tj}$ to facilitate the posterior computation so that $f_{tj}\sim N(m_{tj},s_{tj}^2)$.  
In Section~\ref{sec:real}, we describe how to obtain the means and variances of these normal densities for each agent.

For predicting the multiple time series count data, as the weekly number of COVID-19 inpatients of the prefectures of Japan, a multivariate BPS would improve the predictive performance over a univariate BPS \citep{McAlinn20}. 
However, the specification and implementation of multivariate count time series models are generally cumbersome. 
Therefore, the choice of feasible multivariate models for agents and BPS that can be employed are quite limited. 
Based on the observation that our data exhibit a group of time series with similar dynamics, the proposed model introduced below partitions them into clusters based on the finite mixture of BPS. 
Through the clustering structure, we can leverage the shared information  without resorting to a complex multivariate count model and also reduce the number of parameters to be estimated. 

\subsection{Mixture of Bayesian predictive syntheses}\label{sec:mbps}
Now consider clustering and predicting $n$ univariate time series count data denoted by $y_{it}, i=1,\dots,n,\ t=1,2,\dots$, such as the number of inpatients in $i$th prefecture on $t$th day. 
To predict $y_{it}$, an analyst, again, elicits $J$ models with the predictive density denoted by $h_{itj}(f_{itj}), j=1,\dots,J$. 
The collection of $J$ predictive densities form an information set denoted by $\mathcal{H}_{it}=(h_{it1}(\cdot),\dots,h_{itJ}(\cdot))$. 
The mixture of Bayesian predictive syntheses (MBPS) considers the following synthesised predictive distribution given by
\begin{equation}\label{eqn:mbps}
p(y_{it}|\bPhi_t,\mathcal{H}_{it})
=\int\left[\sum_{k=1}^K\pi_k\alpha(y_{it}|\f_{it},\bPhi_{tk})\right]\left[\prod_{j=1}^Jh_{itj}(f_{itj})\right]d \f_{it},
\end{equation}
where $\sum_{k=1}^K\pi_k\alpha(y_{it}|\f_{it},\bPhi_{tk})$ is the mixture synthesis function whose $k$th component is denoted by $\alpha(\cdot|\f_{it},\bPhi_{tk})$, $\bPhi_{tk}$ denotes the parameter  of $k$th component including the vector of synthesis weights $\bthe_{tk}$, $\pi_k$ is the component weight for $k$th synthesis function with $\sum_{k=1}^K\pi_k=1$, and $\f_{it} = (f_{it1},\dots,f_{itJ})'$ denotes the vector of the latent factors.
With this choice of the synthesis function, the synthesised predictive distribution \eqref{eqn:mbps} is rewritten as
\begin{eqnarray}
p(y_{it}|\bPhi_t,\mathcal{H}_{i,1:t})
&=&\sum_{k=1}^K\pi_k\int\alpha(y_{it}|\f_{it},\bPhi_{tk})\left[\prod_{j=1}^Jh_{itj}(f_{itj})\right]d \f_{it}\nonumber\\
&=&\sum_{k=1}^K\pi_kp_k(y_{it}|\bPhi_{tk},\mathcal{H}_{i,1:t}). 
\label{eqn:mbps2}
\end{eqnarray}
Therefore, MBPS is the finite mixture of synthesised predictive distributions with the synthesis function in 
 $k$th component   given by
\[
p_k(y_{it}|\bPhi_{tk},\mathcal{H}_{i,1:t})=\int\alpha(y_{it}|\f_{it},\bPhi_{tk})\left[\prod_{j=1}^Jh_{itj}(f_{itj})\right]d \f_{it}.
\]

For two time series $i$ and $i'$ in the same cluster $k$, the mixture synthesis function places the same weight $\theta_{tjk}$ on $j$th model whose predictions are generally different between $i$ and $i'$. 
The weights $\bthe_{tk}$ are estimated from the set of time series belonging to the same cluster. 
In this way, MBPS leverages the shared information across time series based on the similarity in how each agent model contributes to the prediction.

The advantages of the proposed MBPS approach are as follows. 
Firstly, MBPS clusters the time series \textit{after} synthesising the individual predictions from $J$ agent models as in \eqref{eqn:mbps}, thus clustering the synthesised predictive distributions. 
Alternatively, one could cluster the $n$ time series within a subset of the agent models. 
However, this approach may be cumbersome since it requires determining the number of clusters for each agent clustering model. 

Secondly, MBPS can reduce the number of parameters to be estimated. 
An alternative BPS approach that does not consider clustering would treat $(y_{1t},\dots,y_{nt})$ as the multivariate time series as in \cite{McAlinn20}. 
In this case, the dimension of the weights is $n\times (J+1)$ for each $t$. 
On the other hand, in MBPS, the dimension of the weights is $K\times (J+1)$ for each $t$ with $K$  much smaller than $n$.

Related to the second point, each agent model fed into MBPS can be univariate. 
Then, these individual univariate predictions are combined through the mixture of synthesised predictive distributions. 
Since the construction and implementation of multivariate count models are generally cumbersome and univariate models are abundant, one can easily choose and fit a set of univariate agent models and feed them into MBPS. 

The model specification for each component is as follows. 
Since $y_{it}$ is the count data, we use the Poisson probability mass function for $\alpha$. 
By introducing the cluster assignment indicator denoted by $z_{i}\in\left\{1,\dots,K\right\}$, $i=1,\dots,n$, 
we have the hierarchical form for MBPS given by
\begin{equation}\label{eqn:Po-synthesis}
\begin{split}
&y_{it}|z_i=k,\bthe_{tk},\f_{it}\sim Poi(\exp(\bthe_{tk}'\F_{it})),\\
&\Pr(z_i=k)=\pi_k,\quad k=1,\dots,K,
\end{split}
\end{equation}
where $Poi(\lambda)$ denotes the Poisson distribution with the mean $\lambda$ and $\F_{it} = (1,\f_{it}')'$. 
Furthermore, the sparse Dirichlet prior distribution $Dir(a_0,\dots,a_0)$ with $a_0<1$ is assumed for $\bpi=(\pi_1,\dots,\pi_K)$ so that the similar components are merged and the weights for the redundant components are shrunk towards zero \citep{RM11}. 
As the default choice, we thus use $a_0=0.01$. 
Under this sparsity-inducing prior, we set $K=n$ in the real data analysis. 
As described in the previous subsection, by constructing the predictive distribution on the log scale, the normal densities are used for $h_{itj}$ such that $f_{itj}\sim N(m_{itj},s^2_{itj})$. 
As in \cite{MW19}, $\bthe_{tk}$ is modelled as the normal random walk model: 
\begin{equation}\label{eqn:theta}
\bthe_{tk} = \bthe_{t-1,k} + \e_{tk},\quad \e_{tk}\sim N(\zero,\bSig_{tk}),\quad k=1,\dots,K. 
\end{equation}

The model is estimated by the Markov chain Monte Carlo (MCMC) algorithm. 
Specifically, we employ the approach of \cite{HIS21} to approximate the probability mass function of the Poisson distribution with that of the negative binomial distribution with a large dispersion parameter. 
Then the P\'olya-gamma augmentation of \cite{POLSON13} is applied to obtain the conditionally linear Gaussian state space model to which the forward filtering and backward sampling algorithm \citep{FS94,WH97} can be implemented. 
We introduce the common discount factors to facilitate the MCMC sampling through the DLM updating scheme. 
The details are provided in Appendix.

\subsection{Introducing heterogeneity in the intercept}
The time series within a cluster of MBPS \eqref{eqn:mbps} have the same contribution from the agent models and the same level of inadequacy of their predictive performance, which is expressed by the intercept. 
However, it would be more helpful to introduce a little flexibility into the model by allowing the intercept to vary within a cluster. 
The extended model with the heterogeneity in the intercept is given by
\begin{equation}\label{eqn:mbps3}
    \begin{split}
    &y_{it}|z_i=k,\bthe_{tk},{\f}_{it},u_{it}\sim Poi(\exp(\bthe_{tk}'\F_{it}+u_{it})),\\
    &u_{it}|z_i=k\sim N(0,\tau_{tk}^2),\\
    &\tau_{tk}^2=\frac{\beta_\tau}{\gamma_t}\tau^2_{t-1,k},\quad \gamma_{t}\sim Beta\left(\frac{\beta_\tau n_{t-1}}{2},\frac{(1-\beta_\tau)n_{t-1}}{2}\right).\\
    \end{split}
\end{equation}
This model can be regarded as an intermediate model between the univariate BPS and MBPS \eqref{eqn:Po-synthesis}. 
The additional random intercept $u_{itk}$ captures the deviation from the cluster-specific constant $\theta_{tk0}$. 
Its variance $\tau_{tk}^2$ follows the beta-gamma random walk volatility model with the discount factor $\beta_\tau\in(0,1]$ \citep[see, for example,][]{WH97,PW10}. 
We use the same Gaussian random walk model for  $\bthe_{tk}$  as in \eqref{eqn:theta}.
Details of the MCMC algorithm are provided in Appendix.

\section{Analysis of COVID-19 inpatients and isolated cases}
\label{sec:real}
Here, the proposed MBPS methodology is demonstrated using the weekly numbers of COVID-19 inpatients in the prefectures of Japan and daily numbers of  COVID-19 isolated cases in the metropolitan autonomous governments of Korea.

\subsection{Agent models}
We consider the following four agent models ($J=4$): Poisson dynamic generalised linear model (DGLM), Poisson generalised additive model (GAM), Poisson integer autoregressive model (INAR) and Power-weighted Poisson model with the susceptible-infected-hospitalised-recovered compartment model (SIHR). 

For DGLM, we assume $y_{it}\sim Poi(\exp(\x_{it}'\bbe_{it}))$ for $i=1,\dots,n$ where $\x_{it}=(1,\tilde{I}_{it}, \tilde{I}_{it}^2)'$ and $\tilde{I}_{it}$ is the log of the seven days lag of the 14 days moving average of the number of infected individuals. 
The time varying regression coefficient $\bbe_{it}$ is assumed to follow the random walk process, $\bbe_{it}=\bbe_{i,t-1}+\u_{it}$ where $\u_{it}$ has the mean $\zero$ and covariance $\W_{it}$. 
The model uses the standard DGLM updating scheme with the discount factor for $\W_{it}$ \citep{WHM85, berry2020bayesian}.

For GAM, we assume $y_{it}\sim Poi(\lambda_{it})$, $\lambda_{it}=\exp(\beta_0+s_{1i}(\tilde{I}_{it})+s_{2i}(t))$ for $i=1,\dots,n$ where $\beta_0$ is the constant term, and $s_{1i}(\tilde{I}_{it})$ and $s_{2i}(t)$ are the smooth functions of $\tilde{I}_{it}$ and $t$, respectively, modelled through the smoothing spline. 
This model is implemented using the R package \texttt{gam}. 

For INAR, we assume $y_{it}|y_{i,t-1}\sim Poi(\exp(\x_{it}'\bbe_i + \gamma_i y_{i,t-1}))$  where $\x_{it}$ is the same as above. 
This model is implemented using the R package \texttt{tscount}. 

Finally, the SIHR epidemic compartment model is considered. 
The SIHR model is chosen because it is reasonably simple and includes a compartment for hospitalisation or isolation. 
The power-weighted approach of \cite{MJ15} discounts the past Poisson observations, which is regarded as a generalisation of the rolling window approach. 
Then, the Poisson means are modelled as the solution of the SIHR epidemic compartment model, obtained by the R package \texttt{deSolve}. 
Specifically, the power-weighted likelihood function for the Poisson observations $y_{i1},\dots,y_{it}$ is given by
\[
p(y_{i,{1:t}}|\lambda_{it})=\prod_{s=1}^{t}\left[\frac{\lambda_{it}^{y_{is}}\exp(-\lambda_{it})}{y_{is}!}\right]^{a_s}, \quad t=1,2,\dots,
\]
where $\lambda_{it}$ is the mean parameter of the Poisson distribution and   $a_s\in[0,1]$ is the discount factor. 
In this approach, the Poisson observations $y_{i1},\dots,y_{it}$ have the common parameter $\lambda_{it}$, but the amount of their information used for the estimation of $\lambda_{it}$ and prediction of $y_{i,t+1}$ is controlled by the discount factors $a_1,\dots,a_t$.
In this paper, it is defined as $a_s=\delta_{SIHR}^{t-s}$ for $s=1\dots,t$ following \cite{FPC20}. 
When $\delta_{SIHR}$ is different from one, the information of the observations further away from the time $t$ is more discounted. 

To construct $h_{itj}$, at the time $t-1$, we calculate the mean and variance of the predictive distribution individually for each agent, $m_{itj}$ and $s_{itj}^2$, given the information $y_{i,1:{t-1}}$ and $\x_{it}$, for the models with covariates. 
These quantities are calculated repeatedly as new information arrives. 
For DGLM, they can be obtained directly from the DGLM updating scheme because the linear predictor is also on the log scale. 
For the remaining three models, we use the parametric bootstrap to compute the predictive means and variances on the log scale. 
Then to predict $y_{it}$, one draws $f_{itj}$ from $N(m_{itj},s_{itj}^2)$ for $j=1,\dots,J$, then given $\f_{it}$, $\bthe_{t z_i}$ and $z_i$, we draw $y_{it}$ from $Poi(\exp(\bthe_{t z_i}'\F_{it}))$. 

In addition to the four agent models, as suggested by one of the reviewers, we also consider the finite mixture of Poisson regression models, denoted by FMPR.
This model is based on \cite{ZZ}, which is also reviewed by \cite{FS11}. 
While FMPR is not included in BPS as an agent, it would be interesting to compare the proposed BPS models with a model from a different class. 
The specification of FMPR is similar to that of our MBPS, but instead of latent factors from the agent models, it includes the observable covariates: 
\[
    \begin{split}
    y_{it}|z_i=k&\sim Poi(\exp(\x_{it}'\bbe_{k})),\\
    \Pr(z_i=k)&=\pi_k,\quad k=1,\dots,K
    \end{split}
\]
where $\x_{it}=(1,\tilde{I}_{it},\tilde{I}_{it}^2,y_{i,t-1})'$. 
Assuming the prior distribution $\beta_k\sim N(\zero,100\I)$, the model is estimated using the MCMC.

\subsection{Performance measures}
We compare the performance of the agent models, FMPR, and the proposed BPS model for univariate counts and three MBPS models. 
Hereafter, MBPS denotes the model with \eqref{eqn:Po-synthesis} and MBPSH denotes MBPS with the heterogeneous intercepts in \eqref{eqn:mbps3}. 
The univariate BPS model is denoted by BPS when there is no confusion. 
The performance of the models is compared based on the following: 
the cumulative absolute prediction errors (CAPE) given by
\[
\text{CAPE}_{it}^s = \sum_{t^*=T}^{t}|y_{i,t^*+s}-\hat{y}_{i,t^*+s}|, \quad
\]
the log predictive density ratio (LPDR) given by
\[
\text{LPDR}_{it}^s=\log \frac{p_j(y_{i,t+s}|y_{i,1:t})}{p_\text{MBPS}(y_{i,t+s}|y_{i,1:t})}, \quad t=T,\dots,T+T^*-s,
\]
where $p_\text{MBPS}(y_{t+s}|y_{1:t})$ and $p_j(y_{t+s}|y_{1:t})$ are the $s$-step-ahead predictive density of MBPS  and other models, respectively, and 
the coverage of 95\% prediction intervals given by
\[
\text{Coverage}^s = \frac{1}{n(T^*-s)}\sum_{i=1}^n\sum_{t^*=T}^{T+T^*-s}\left\{L_{i,t^*+s}<y_{i,t^*+s}<U_{i,t^*+s}\right\}
\]
The LPDR for $j$th model below zero implies the superiority of the proposed MBPS, and that above implies the superiority of $j$th model over MBPS. 
We also consider the total CAPE defined by $\sum_{i=1}^{n}\text{CAPE}_{it}^s$ and total LPDR defined by $\sum_{i=1}^n\text{LPDR}_{it}^s$. 

\subsection{Analysis of latent factors}
Following \cite{MW19}, we compute the following measures to investigate the inter-dependency and redundancy in the agent predictions based on the latent factors inferred from the observed data. 
For each $t$, we compute the Monte Carlo (MC)-empirical $R^2$ for the latent factor for $j$th agent, $f_{itj}$, given the factors for all the other agents, $f_{itj'},\ j'\neq j$ from the MCMC output. 
This $R^2$ measures the redundancy of $j$th agent in the context of all $J$ agents for $j=1,\dots,J$. 
We also compute the MC-empirical $R^2$ for each pair of $j$ and $j'$, $j\neq j'$, to measure the pairwise dependence. 
It is called the paired MC-empirical $R^2$.

\subsection{Cluster determination}
The MCMC algorithm described in Appendix provides the posterior sample of the cluster assignment, denoted by $\z^{(1)},\dots,\z^{(L)}$ where $\z^{(\ell)}=(z_1^{(\ell)},\dots,z_N^{(\ell)})$, $\ell=1,\dots,L$. 
Following \cite{NC14} and \cite{LIN19}, we select the most representative draw $\z^{(\ell^*)}$ from the MCMC output. 
To do so, define the $N\times N$ co-clustering matrix $\Z^{(\ell)}$ for $\ell=1,\dots,L$ whose $(i,j)$ entry is equal to one if $z_i^{(\ell)}=z_j^{(\ell)}$ and is equal to zero otherwise. 
We also define the mean co-clustering matrix $\hat{\Z}=L^{-1}\sum_{\ell=1}^L\Z^{(\ell)}$ which summarises the entire MCMC output. 
Then we select the $\ell^*$th draw $\z^{(\ell^*)}$ as the final clustering for which the Frobenius norm from the mean co-clustering matrix is minimised: $\ell^*=\arg\min_\ell \|\Z^{(\ell)}-\hat{\Z}\|_F$.

\subsection{Analysis of weekly numbers of inpatients in Japan}
The agent models are estimated using the first 50 weeks up to 14 April 2021. 
The BPS models are first run using the data from 21 April  2021 to 30 March 2022 ($T=50$ weeks). 
Then, by expanding the window of the past data, the one-step-ahead predictions are obtained for both agent and BPS models from 6 April 2022 to 23 November 2022 ($T^*=34$ weeks). 
All the discount factors for the agent and BPS models are set to $0.95$ to account for the fast changes in the data.

\subsubsection{Predictive performance}
\begin{figure}[H]
\includegraphics[scale=0.28]{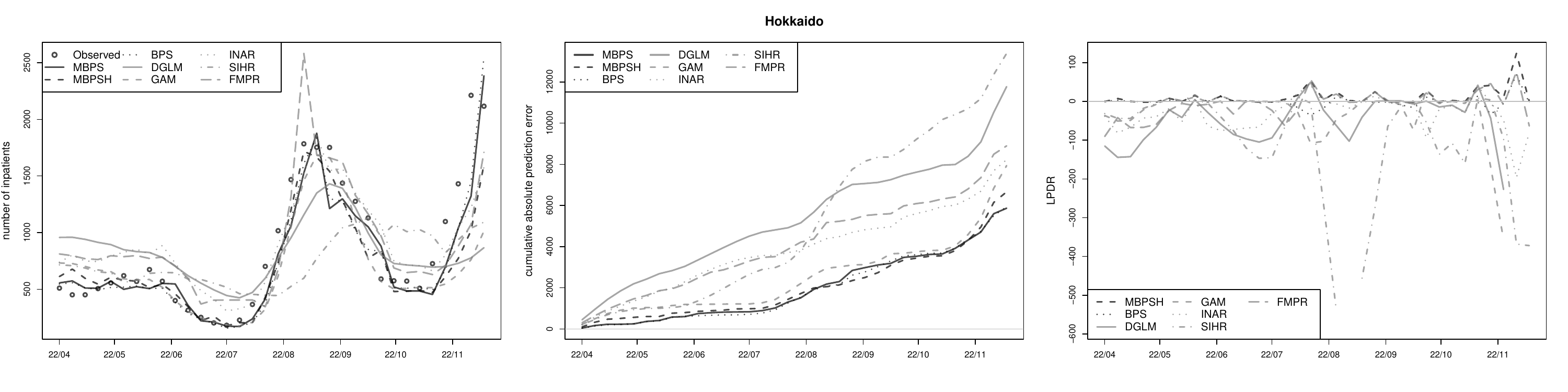}
\includegraphics[scale=0.28]{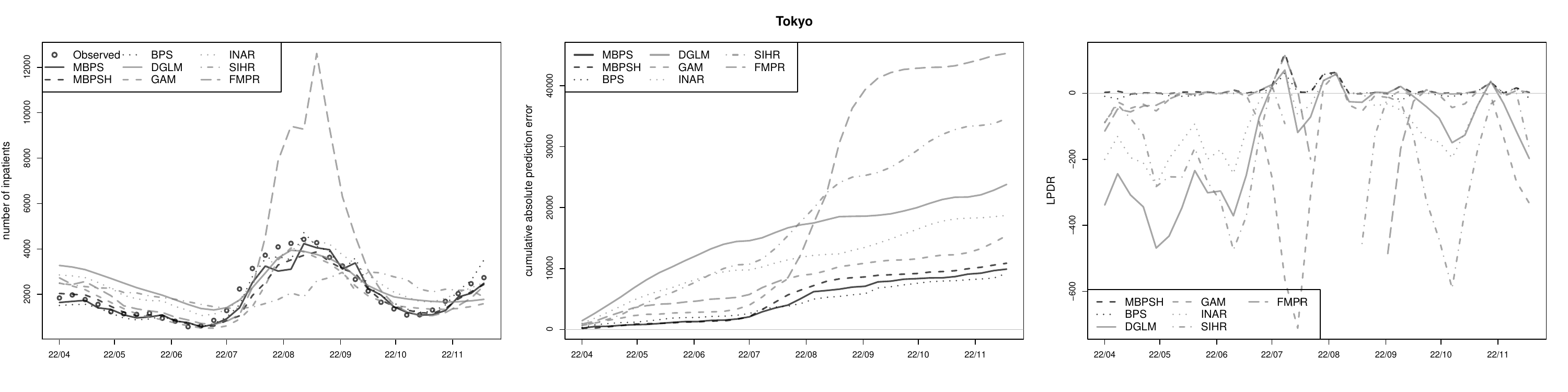}
\includegraphics[scale=0.28]{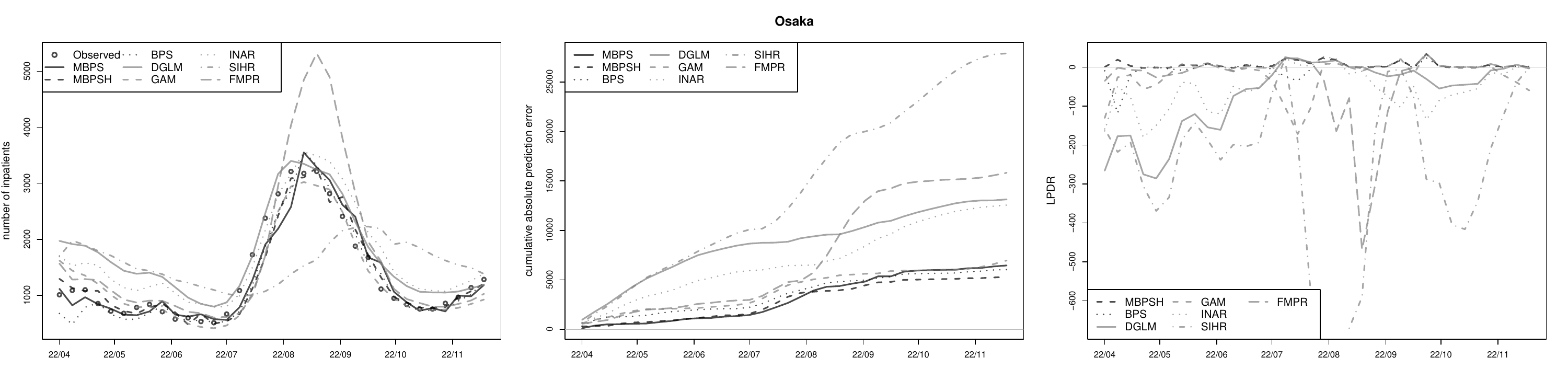}
\includegraphics[scale=0.28]{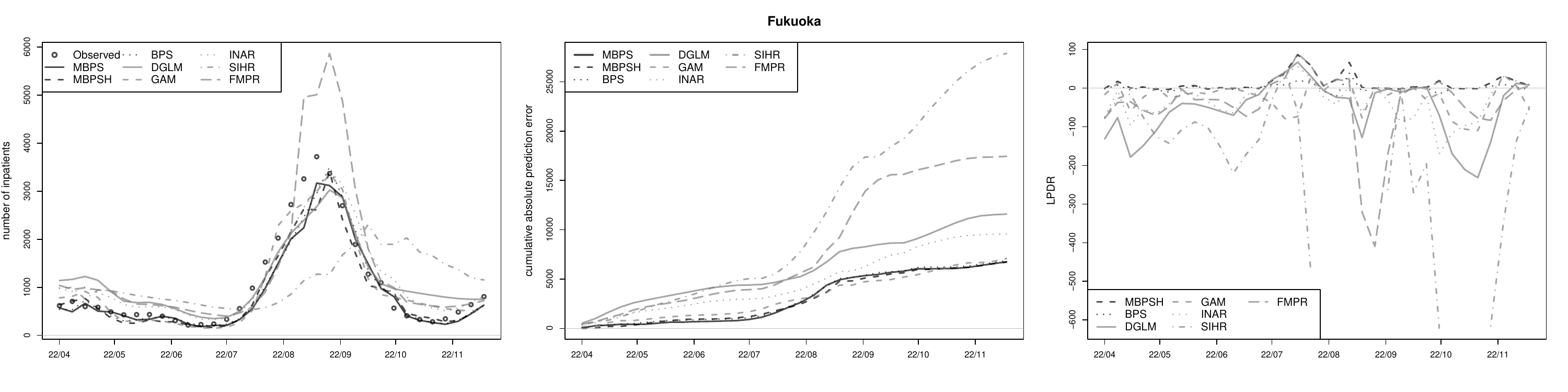}
\caption{One-step-ahead prediction results (left), CAPE (middle) and LPDR (right) for the selected prefectures}
\label{fig:res_pref}
\end{figure}

Figure~\ref{fig:res_pref} presents the prediction results under the agent models and MBPS, CAPE and LPDR for the selected prefectures, Hokkaido, Tokyo, Osaka and Fukuoka. 
The left panels of the figure show that the proposed MBPS, MBPSH and BPS perform reasonably well overall, tracking the fast-changing time series. 
The CAPE of these three models are comparable and are smaller than or at least comparable with those of the agent models. 
On the contrary, for the agent models, especially SIHR, there seems to be some lag in turning from a downward trend to an upward trend and vice versa, causing larger prediction errors. 
The predictions under the agent models tend to be larger than the observed data in the downward trend and smaller in the upward trend. 
When the number of inpatients peaked in August 2022, the predicted number of inpatients under FMPR was much larger than the observed number. 
The LDPR for the agent models are negative for most of the prediction period, except for July of 2022, during which the number of inpatients turned into an upward trend towards the peak in August. 
There should not be a concern about this since the predictions under the agent models are generally higher than the observed trend before July, and they happen to become close to or cross the observed data in July, resulting in positive LPDR.  
For some periods, LPDR, especially for SIHR, was not obtained due to the underflow and thus, the curves are discontinuous during those periods. 

\begin{figure}[H]
\includegraphics[scale=0.35]{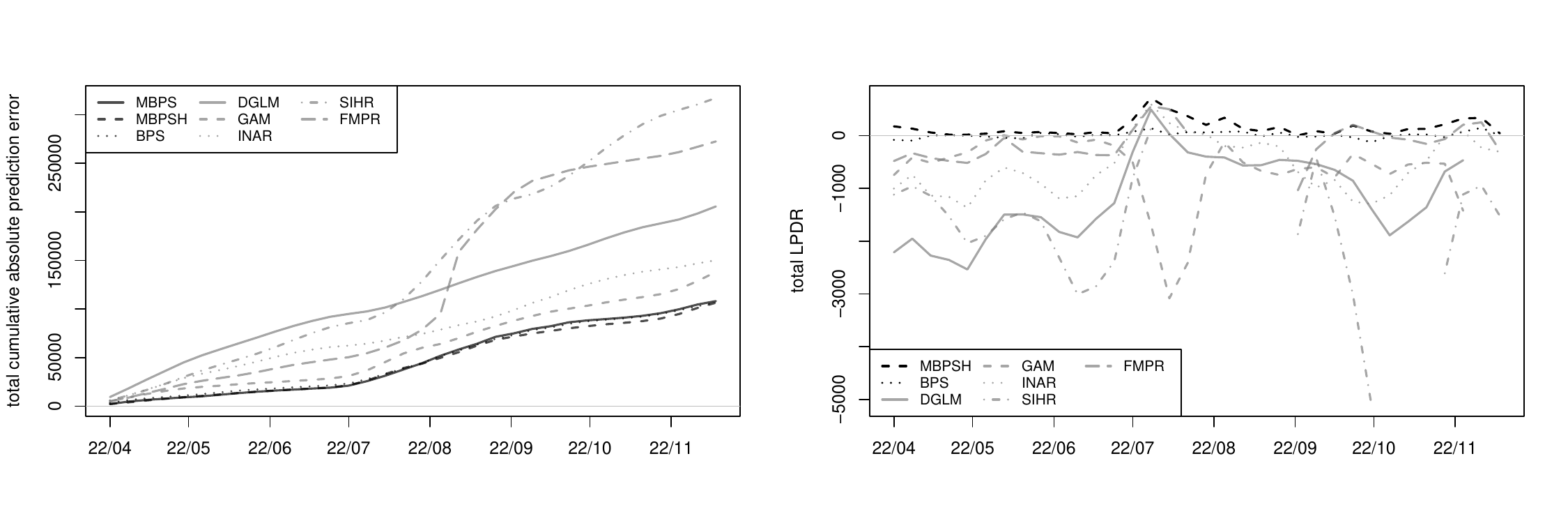}
\caption{One-step-ahead CAPE (left) and LPDR (right) summed over prefectures}
\label{fig:res_total_jp}
\end{figure}

Figure~\ref{fig:res_total_jp} presents the total CAPE and total LPDR, which are the CAPE and LPDR summed over the prefectures, as a summary of the overall performance of the models. 
It is seen that the total CAPE  for MBPS, MBPSH and BPS are the lowest of the nine models, and they are comparable. 
The figure also shows that the total CAPE for SIHR is the largest at the end of the prediction period, followed by FMPR. 
The right panel of the figure shows that LPDR for the agent models are negative for most of the prediction period, implying the superiority of MBPS over these models. 
On the other hand, LPDR for MBPSH is positive throughout the prediction period. 

Table~\ref{tab:cov_jp} presents the coverages of the 95\% prediction intervals for one-step prediction. 
MBPSH resulted in the coverage closest to the target value of $0.95$. 
The other models resulted in undercoverage, and the coverages for the agent models and FMPR are particularly low, but some improvements by MBPS and BPS, which are comparative, over these models are also observed. 
This result highlights the severe difficulty in accurately predicting the number of COVID-19 inpatients. 
Appendix presents the results for the cases where the variances of the factors, $s_{itj}^2$, in the BPS models, are deliberately inflated by five and times. 
However, the table shows little differences between the cases. 
In what follows, we mainly focus on MBPSH, which performs the best of the models considered here.

\begin{table}[H]
\caption{Coverages of 95\% prediction intervals for one-step-ahead prediction for Japanese data under the agents, FMPR and  BPS models}
\label{tab:cov_jp}
\centering
\begin{tabular}{crrrrrrrr}\hline
  MBPS &       MBPSH & BPS &   DGLM  &     GAM   &   INAR  &     SIHR & FMPR\\\hline
0.556 &0.928& 0.583& 0.158& 0.122& 0.236& 0.036& 0.287\\
\hline
\end{tabular}
\end{table}

\subsubsection{Analysis of latent factors}
\begin{figure}[H]
 \includegraphics[scale=0.28]{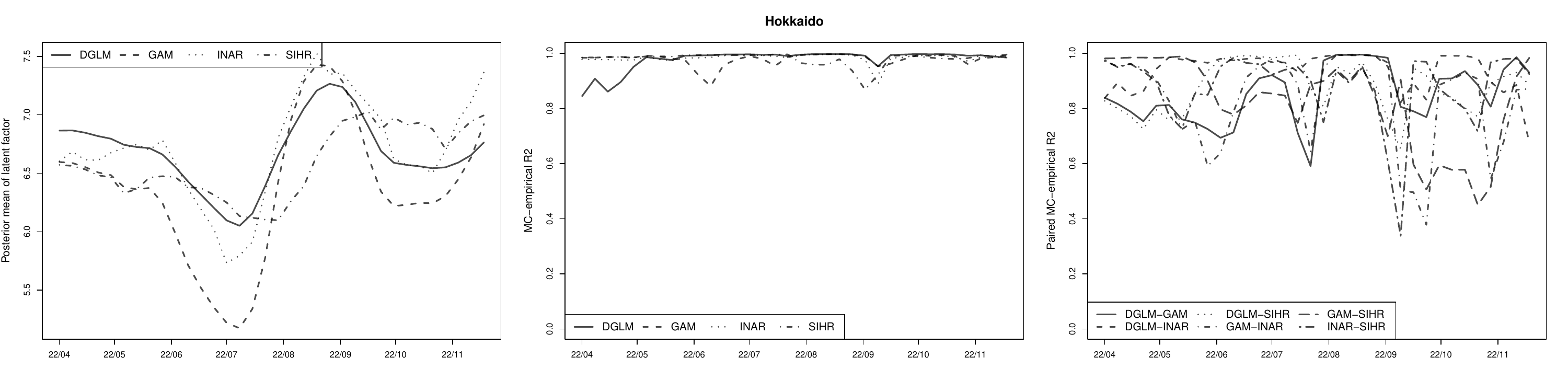}
\includegraphics[scale=0.28]{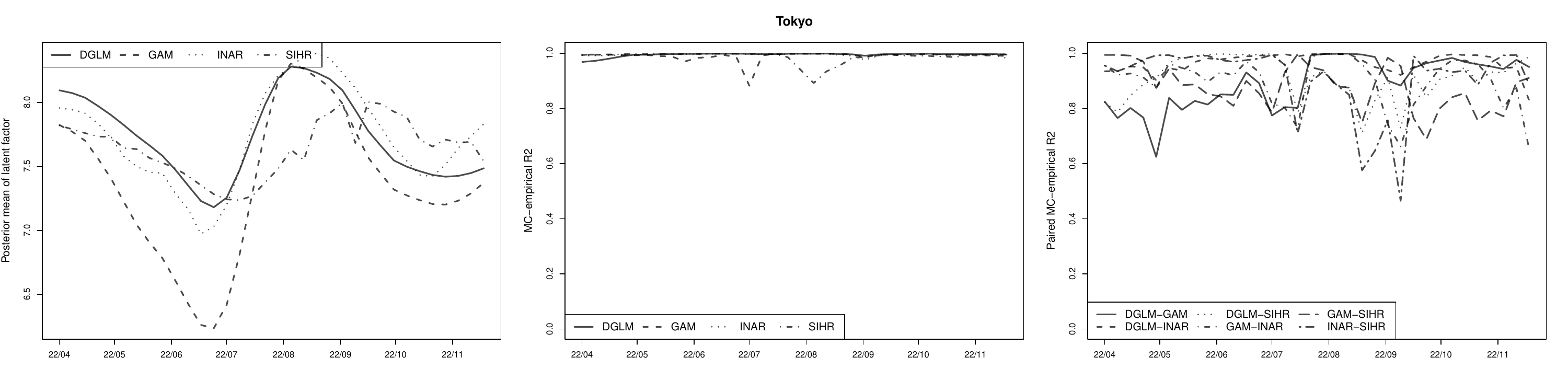}
\includegraphics[scale=0.28]{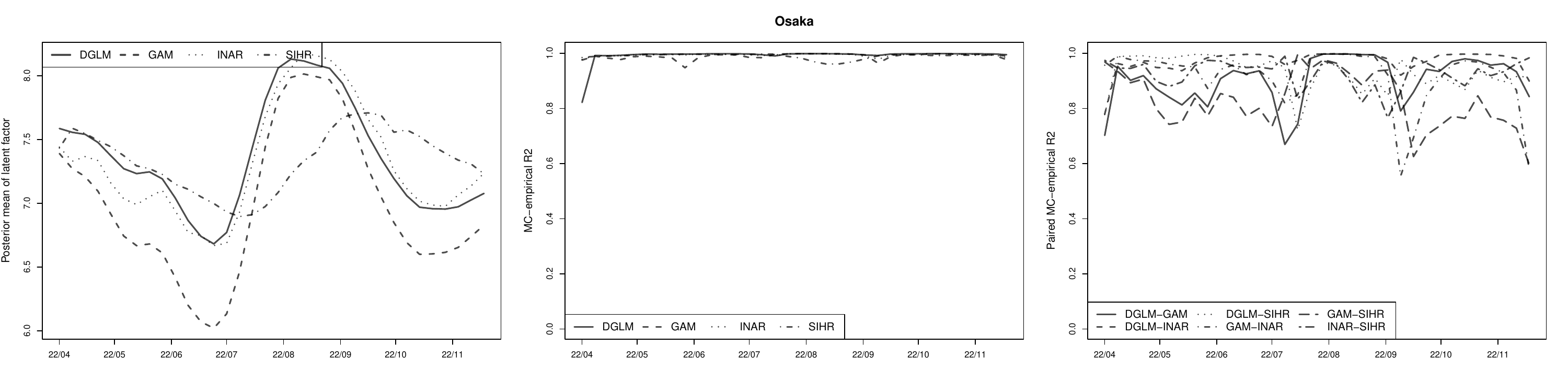}
\includegraphics[scale=0.28]{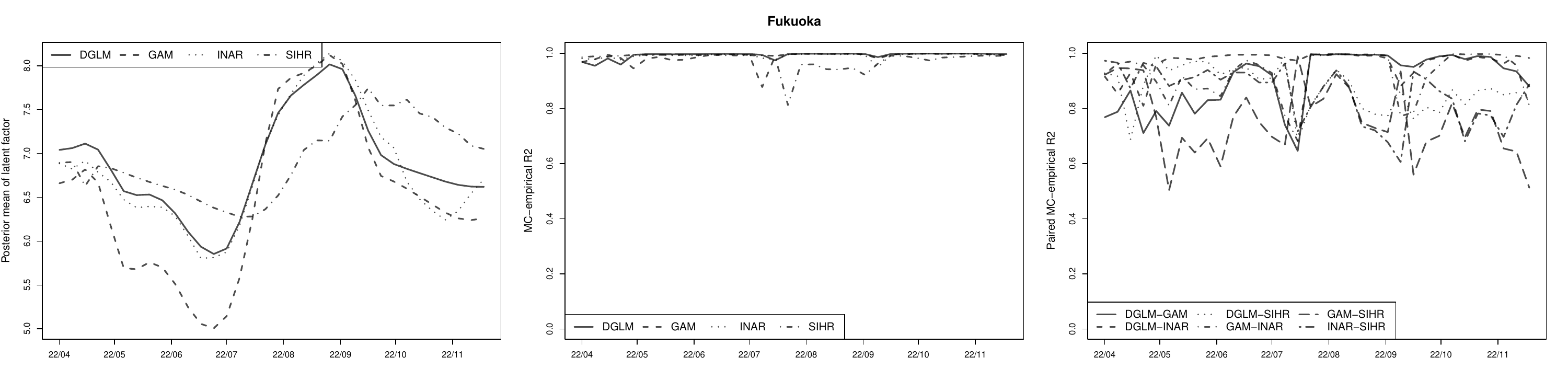}
\caption{Posterior means of the latent factors $f_{itj}$ (left), MC-empirical $R^2$ (middle) and paired MC-empirical $R^2$ (right) under MBPSH for the selected prefectures given all Japanese data}
\label{fig:res_pref_factors}
\end{figure}

Following \cite{MW19}, we have a closer look at the posterior inference on the latent factors of MBPSH given all data up to 23 November 2022. 
Figure~\ref{fig:res_pref_factors} presents the posterior means of the latent factors, MC-empirical $R^2$ and paired MC-empirical $R^2$ for the same selected prefectures as in Figure~\ref{fig:res_pref}. 
The left panels show that the posterior means of the latent factors, especially for DGLM, GAM and INAR, draw similar trajectories, albeit with some differences in amplitudes. 
This is reflected in the middle panels of the figure where the MC-empirical $R^2$ are very close to one. 
It is also seen that the predictability of SIHR, given others, drops during the peak period of August 2022, especially for Fukuoka. 
The paired $R^2$ are generally smaller than $R^2$ but are still above $0.6$ in most of the prediction period for these prefectures. 

\begin{figure}[H]
\includegraphics[scale=0.37]{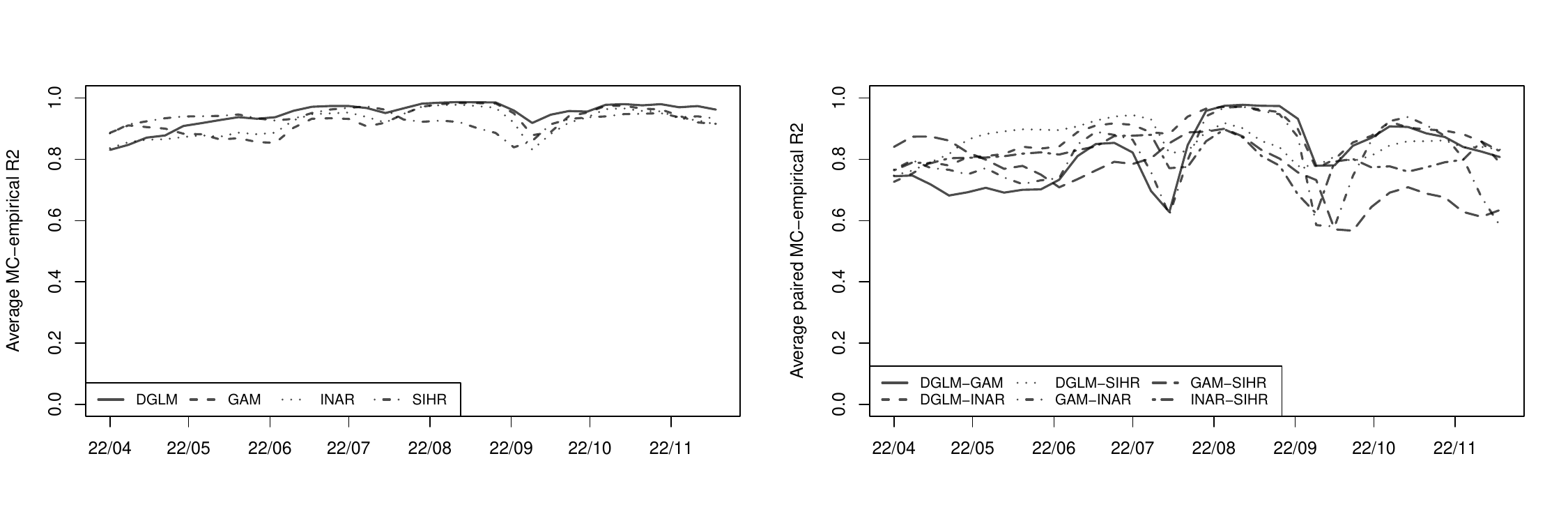}
\caption{MC-empirical $R^2$ (left) and paired MC-empirical $R^2$ (right) averaged over prefectures given all Japanese data
}
\label{fig:res_total_jp_factors}
\end{figure}

The overall pattern in these measures can be more clearly seen by averaging over the prefectures, as presented by Figure~\ref{fig:res_total_jp_factors}. 
The average MC-empirical $R^2$ are generally high, between $0.8$ and $1$, throughout the prediction period, as shown in the left panel. 
The average $R^2$ for SIHR is slightly low compared to the others during the peak period of August 2022. 
It is also seen that the average $R^2$ for all agents dropped in September 2022, indicating some degree of disagreement  among the agents on predicting when the peak has passed. 
The right panel of the figure shows that the average $R^2$ for the pairs DGLM-SIHR, GAM-SIHR and INAR-SIHR are low compared to the rest of the pairs in the peak period.  
It is also seen that the average paired $ R2$ dropped before and after the peak, July 2022 and September 2022.

\subsubsection{Single agent method with different features}
We investigate the cases where the agent models in a BPS model are confined within a single model class but with different features rather than using four different methods.
This is also done in \cite{TM23}. 
More specifically, under MBPSH, we consider the following three cases. 
In the first case, MBPSH uses three DGLM with different set of covariates as agents, denoted by DGLM1 with $\x_{it}=1$, DGLM2 with $\x_{it}=(1,\tilde{I}_{it})'$ and DGLM3 with $\x_{it}=(1,\tilde{I}_{it},\tilde{I}_{it}^2)'$. 
In the second case, MBPSH uses three GAM with the different mean functions as agents, denoted by GAM1 with $\lambda_{it}=\exp(\beta_0+s_{1i}(\tilde{I}_{it}))$, GAM2 with $\lambda_{it}=\exp(\beta_0+s_{2i}(t))$ and GAM3 with $\lambda_{it}=\exp(\beta_0+s_{1i}(\tilde{I}_{it})+s_{2i}(t))$.  
Finally, we consider MBPSH with three INAR with different covariates as agents. 
The setting for the covariates is the same as that for DGLM.
In this case, the corresponding agents are denoted by INAR1, INAR2 and INAR3. 
MBPSH-DGLM, MBPSH-GAM and MBPSH-INAR denote these specifications for MBPSH. 

\begin{figure}[H]
    \centering
    \includegraphics[scale=0.33]{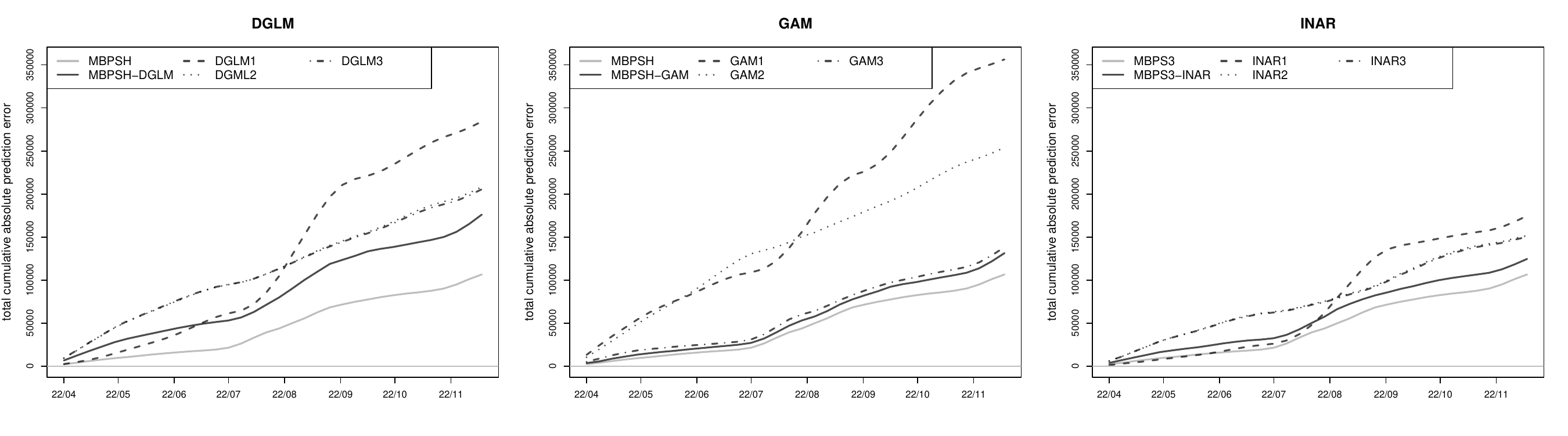}

    \caption{Total CAPE for MBPSH with four different agent methods and MBPSH with the agents in a single model class with different features for Japanese data}
    \label{fig:only_agents_jp}
\end{figure}

Figure~\ref{fig:only_agents_jp} presents the total CAPE in the three cases. 
The figure also plots the total CAPE for MBPSH with four different agent methods shown in grey in Figure~\ref{fig:only_agents_jp}. 
In all cases, the original MBPSH resulted in the lowest total CAPE. 
In each case, the agent model with the smallest set of covariate information resulted in the largest CAPE.
For DGLM and INAR, the results for DGLM2 and DGLM3, as well as those for INAR2 and INAR3, are comparable, implying that the smaller sets of covariates may be sufficient. 
For GAM, MBPSH-GAM is comparable with GAM3. 
The result shown in Figure~\ref{fig:only_agents_jp} suggests that including and combining various possible predictions in a BPS model is more effective than limiting the agent method to a single class. 

\subsubsection{Clustering results}\label{sec:clust_jp}
\begin{figure}[H]
\centering
\includegraphics[scale=0.35]{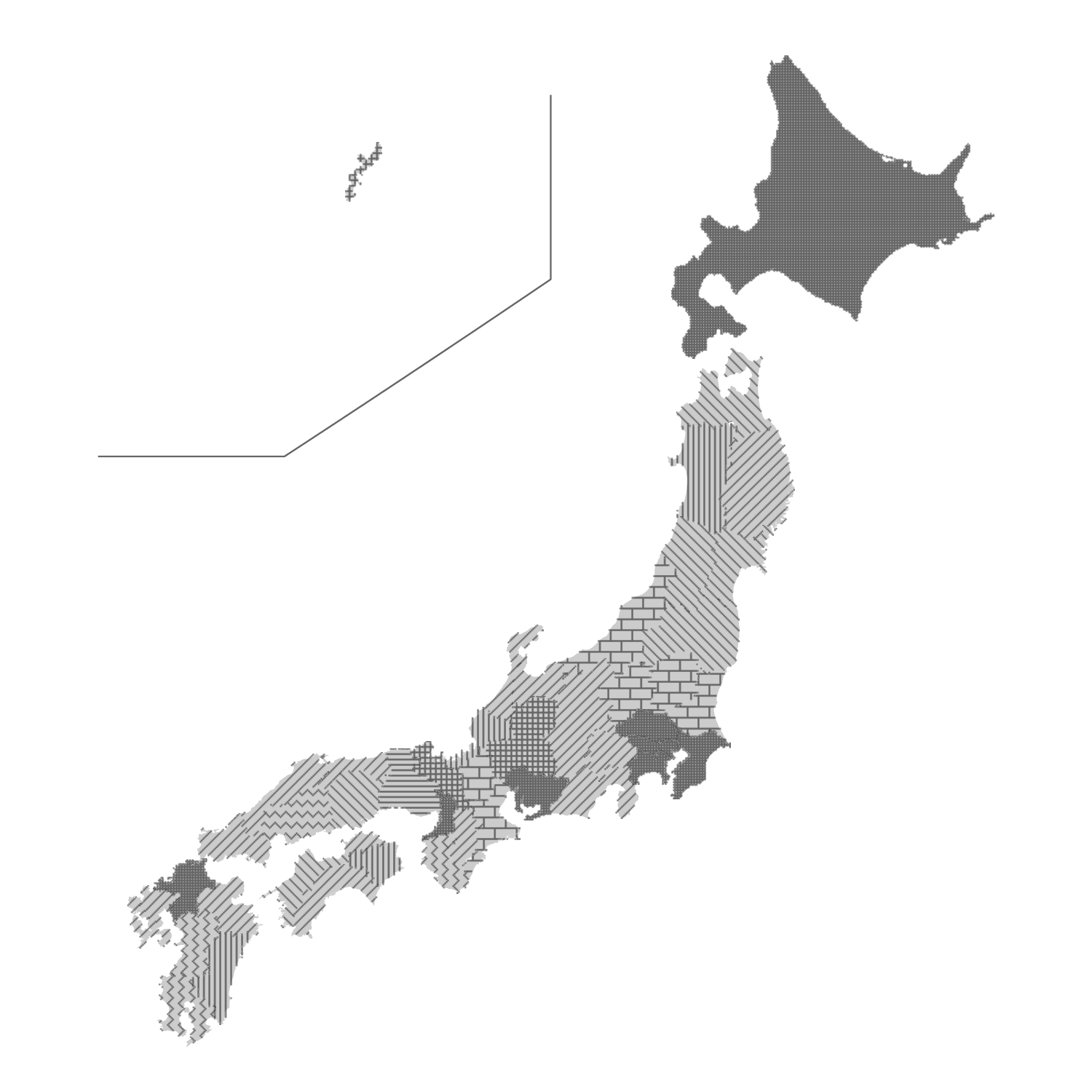}
\includegraphics[scale=0.3]{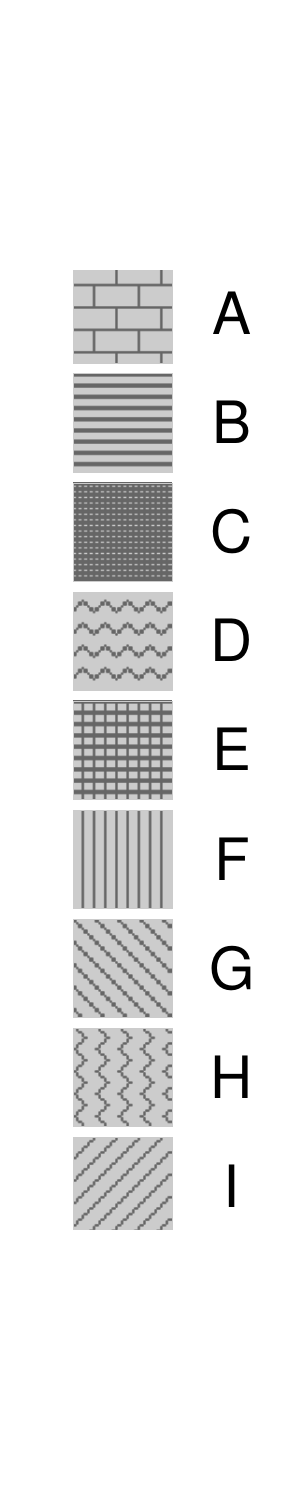}
\includegraphics[scale=0.55]{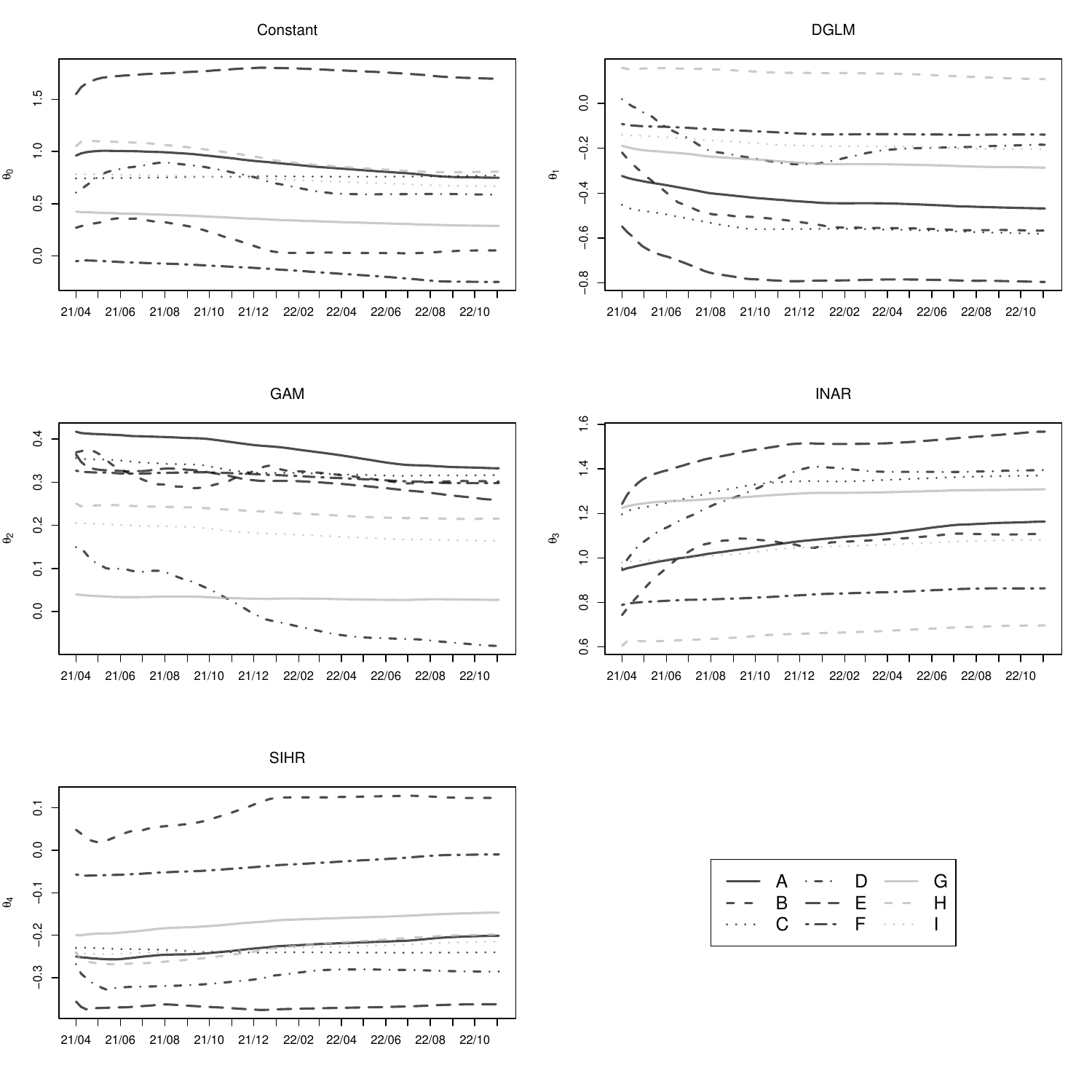}
\caption{Clustered prefectures (top map), posterior means of the synthesis weights under MBPSH (bottom panels) given all Japanese data
}
\label{fig:res_cluster_jp}
\end{figure}

Figure~\ref{fig:res_cluster_jp} presents the clustered prefectures and posterior means of the synthesis weights based on the clustering results under MBPSH using all data up to 23 November 2022. 
Although a geographical interpretation of the clustering result may not be straightforward as MBPS clusters the prefectures based on the contribution of the agent models, we notice that the prefectures with similar magnitudes in the time series happened to form a cluster. 
Cluster~C includes Tokyo, Chiba, Saitama, Kanagawa, which are the neighbours of Tokyo,  Hokkaido, Osaka, Aichi and Fukuoka. 
Cluster~E Gifu, Kyoto and Okinawa. 
The prefectures in these clusters incurred large numbers of inpatients, as shown in the bottom panels of Figure~\ref{fig:jp}. 

In the bottom panels of Figure~\ref{fig:res_cluster_jp}, the posterior means of the synthesis weights exhibit smooth trajectories with different levels for different clusters. 
For the constant term, Cluster~E resulted in the largest values, followed by Clusters~H and A, indicating the difficulty in predicting for those prefectures. 
For DGLM, the weights for most clusters are negative, and Clusters~E resulted in the largest magnitudes. 
GAM has a large predictive contribution in Clusters~A, B, C, E and F. 
On the contrary, the posterior means are close to zero for Cluster~G throughout the data period. 
The weight for Cluster~D turns from positive to negative in November 2021. 
The weights for INAR are positive and generally large for all clusters. 
Cluster~E resulted in the largest posterior means. 
For SIHR, Cluster~B resulted in the posterior means of the largest magnitudes. 
Again, the number of inpatients is challenging to predict, so allowing negative weights in an agent model with poor predictive performance might improve the overall performance of BPS. 

\begin{figure}[H]
\centering
\includegraphics[scale=0.4]{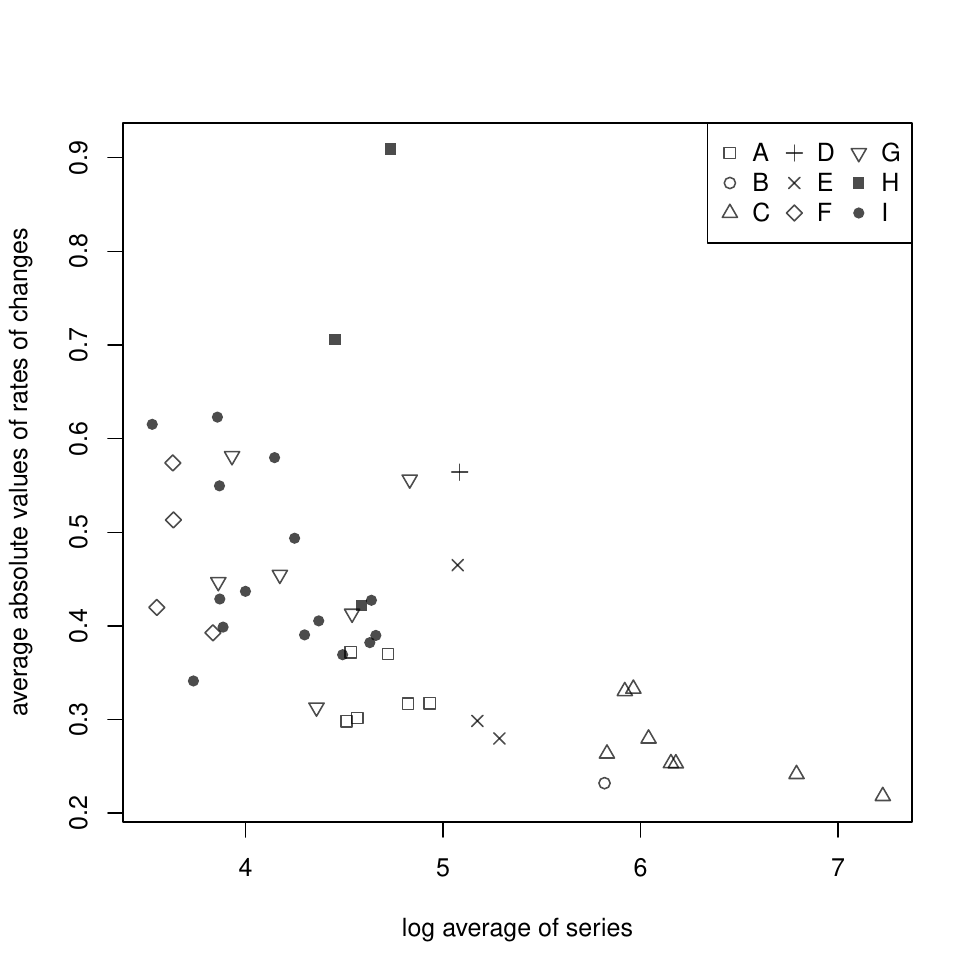}
\caption{Profile of the clustered time series for the Japanese data given all data}
\label{fig:clusters_prof_jp}
\end{figure}

Figure~\ref{fig:clusters_prof_jp} presents the scatter plot of the log averages of the time series, $\log(T^{-1}\sum_{t=1}^Ty_{it})$, and average absolute values of the rates of changes from previous weeks in the time series, $T^{-1}\sum_{t=2}^T|y_{it}-y_{i,t-1}|/y_{i,t-1}$, to gain more insight on the characteristics of the clustered time series in terms of the level and smoothness of the series. 
It is first seen that the locations of the series belonging to the same cluster are similar. 
As described above regarding Figure~\ref{fig:res_cluster_jp}, the levels of the time series appear to be an important factor in the clustering of MBPS. 
It is also seen that the clusters with large numbers of inpatients,  such as Clusters~B and C, located in the bottom right corner of the figure, are also associated with relatively low rates of changes compared to the rest of the clusters. 
Since GAM includes the smooth function of the number of infected individuals and time index, we expected that the clusters with large magnitudes of synthesis weights for GAM, such as Clusters~A, B and C, E, would be associated with low rates of changes in the time series. 
This pattern is partly observed in the figure.

\begin{figure}[H]
\centering
\includegraphics[scale=0.5]{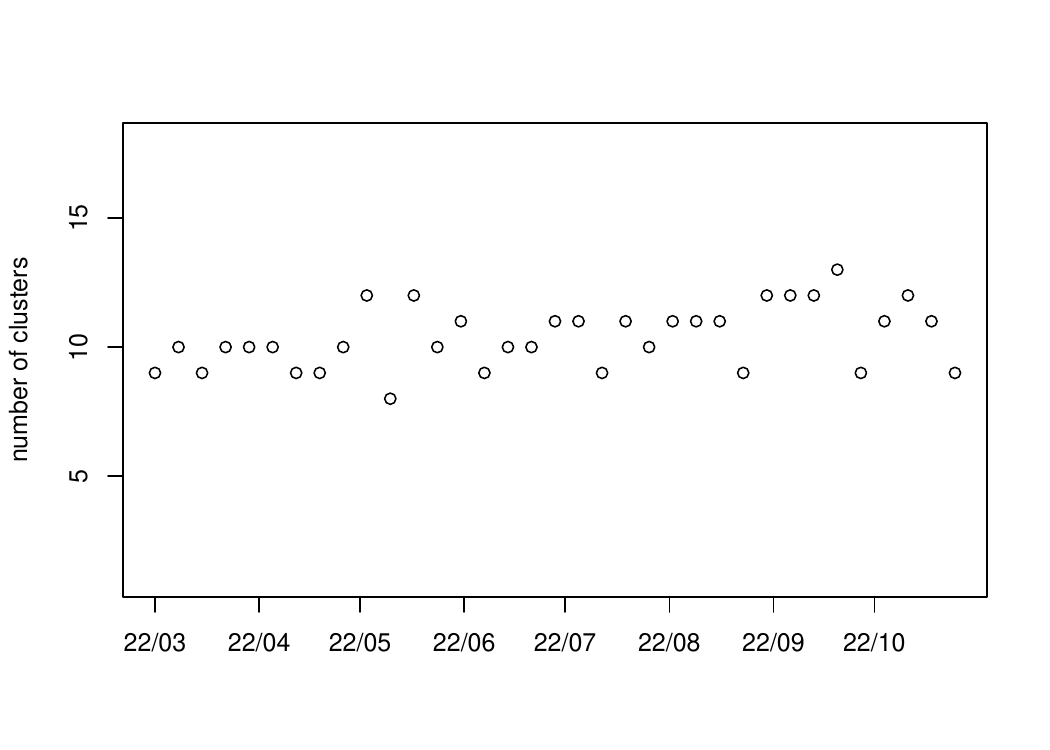}
\caption{Numbers of clusters under MBPSH for the Japanese data}
\label{fig:num_clusters_jp}
\end{figure}

Finally, Figure~\ref{fig:num_clusters_jp} shows the numbers of the determined clusters under MBPSH over the data period. 
The figure shows that the numbers of alive clusters do not vary significantly. 
The average number of clusters is $10.4$.

\subsection{Analysis of daily isolated cases in Korea}
As in the previous analysis, the same four agent models are considered and are first estimated using the first three months of the data up to 31 October 2020. 
MBPS is fit using the data from 1 November 2020 to 31 July 2021 ($T=273$ days) and then by expanding the window of the past data, the one-, three- and seven-step-ahead ($s=1,3,7$) predictions from 1 August  2021 to 30 November 2021, are obtained ($T^*=122$ days).

The multistep-ahead forecasts of MBPS are based on the customised BPS approach proposed by \cite{MW19}. 
Specifically, for predicting $y_{t}$ at time $t-s$, BPS includes 
the $s$-step-ahead predictions of the agent models $h_{i,t-s,j}(f_{itj})$ rather than $h_{itj}(f_{itj})$. 
While the horizon-specific BPS model has to be estimated for each prediction horizon, it directly relates the outputs from the agent models to the horizon $s$ of interest. 
The discount factors for the agent and BPS models are set to $0.99$. 

\subsubsection{Predictive performance}
\begin{figure}[H]
\centering
\includegraphics[scale=0.28]{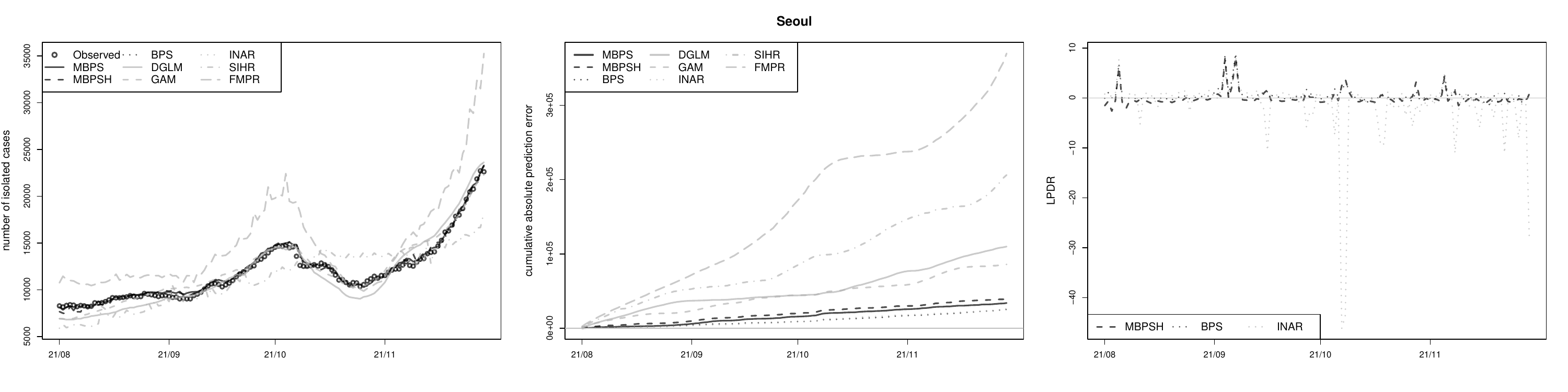}
\includegraphics[scale=0.28]{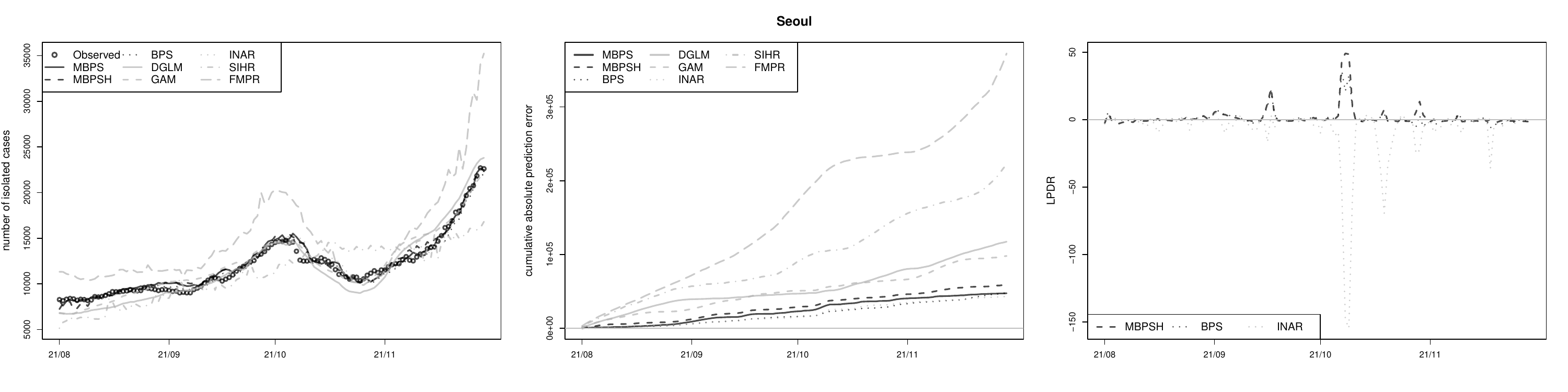}\\
\includegraphics[scale=0.28]{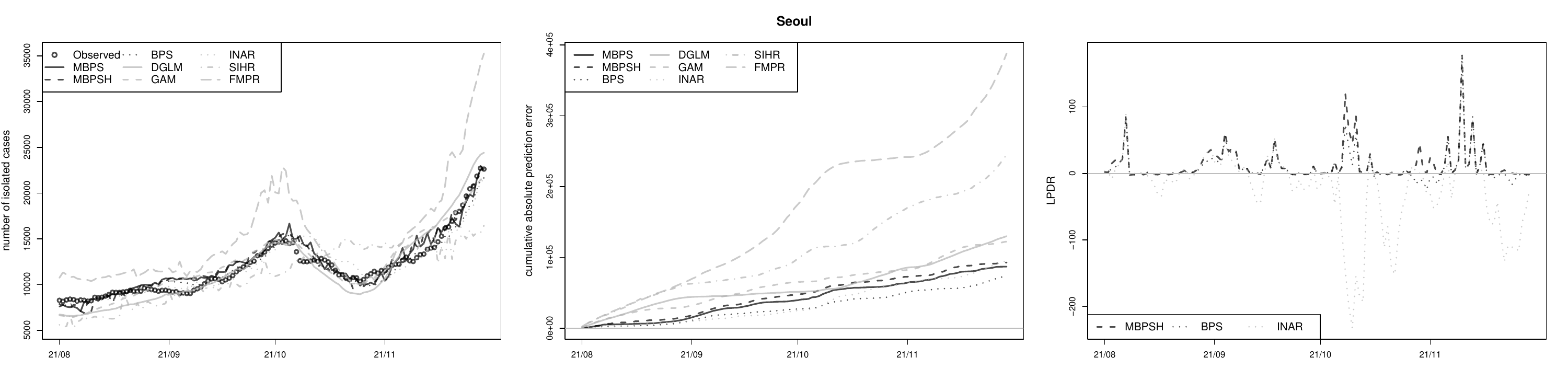}\\
\caption{One-, three- and seven-step-ahead prediction results  (left), CAPE (middle) and LPDR (right) for Seoul }
\label{fig:res_seoul}
\end{figure}

Figure~\ref{fig:res_seoul}  presents the prediction results, CAPE and LPDR  for $s=1,3$ and $7$ under the agent and proposed models for Seoul. 
The LPDR is shown only for MBPSH, BPS and INAR for visibility. 
In the left panels of Figure~\ref{fig:res_seoul}, SIHR, and FMPR clearly fail to predict the number of isolated cases. 
The middle panels of Figure~\ref{fig:res_seoul} show that CAPE for MBPS, MBPSH, BPS and INAR are comparable throughout the prediction period in the cases of $s=1$ and $3$. 
More specifically, CAPE for BPS and INAR are slightly smaller than those for MBPS and MBPSH. 
In the case of $s=7$, BPS resulted in the smallest CAPE, followed by MBPS, INAR and MBPSH. 
In the right panels, LPDR for INAR exhibits occasional downward spikes, particularly around the mild peak in October 2021 and towards the end of the prediction period, especially for $s=7$. 
These indicate the superiority of the BPS models over INAR. 
Between MBPS and MBPSH or BPS, the figure also shows the notable upward spikes in LPDR in October 2021, indicating MBPSH and BPS performed better than MBPS in capturing this mild peak in the number of isolated individuals.

\begin{figure}[H]
\centering
\includegraphics[scale=0.35]{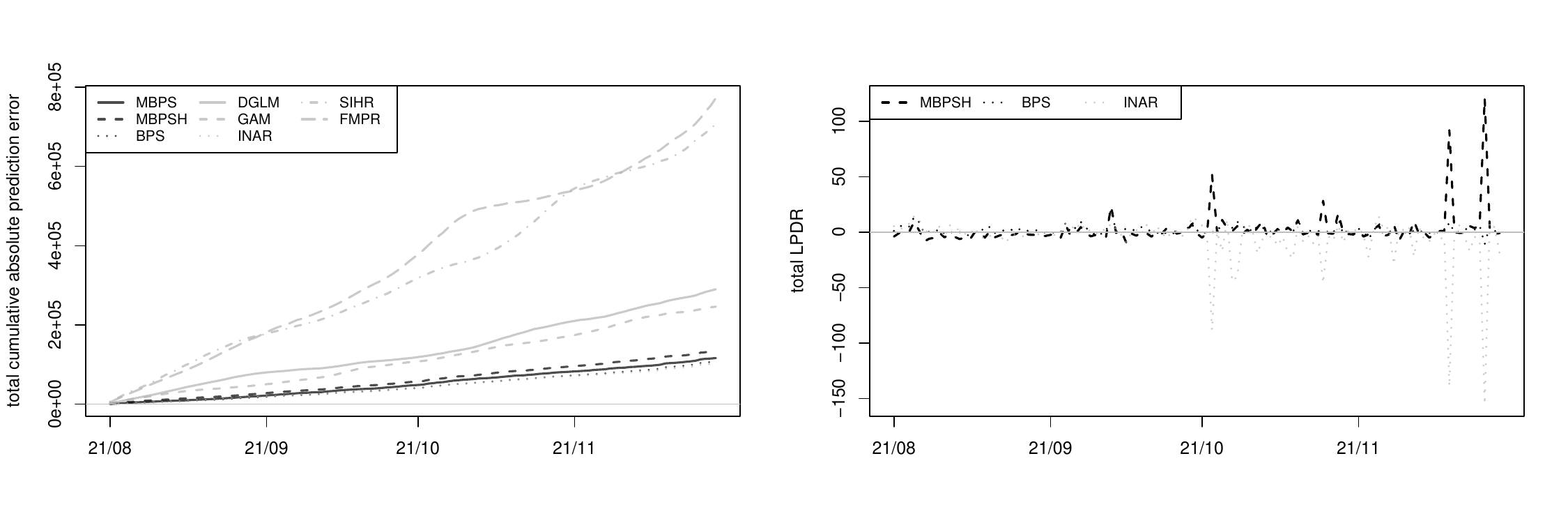}
\includegraphics[scale=0.35]{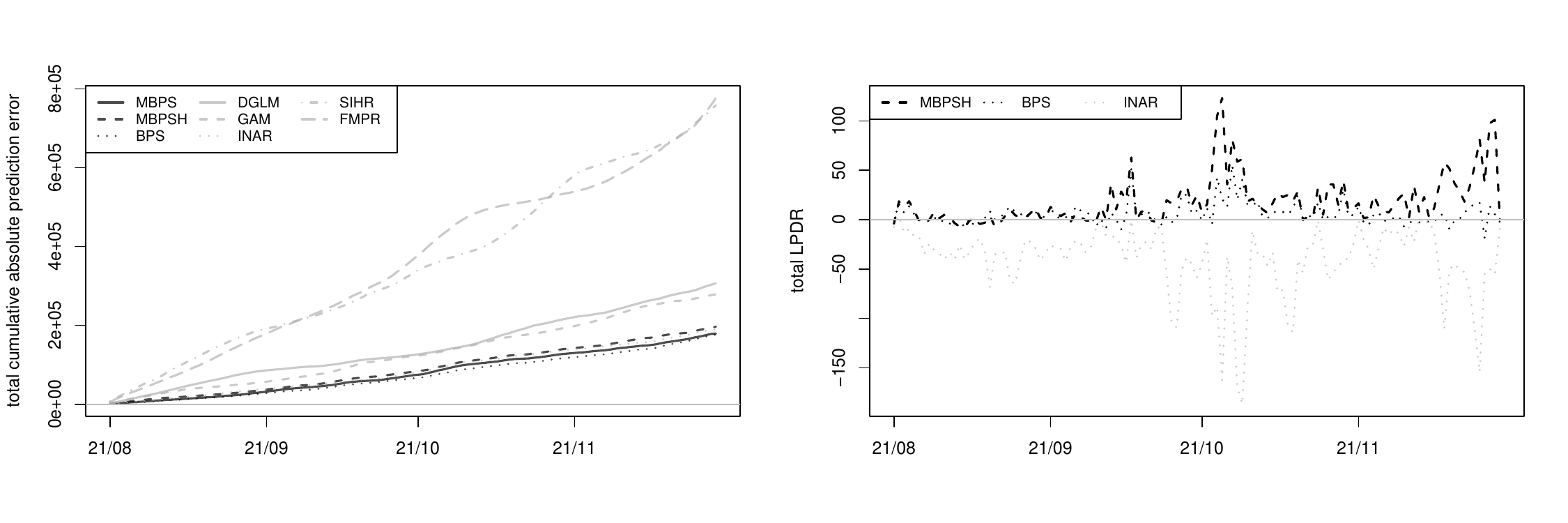}
\includegraphics[scale=0.35]{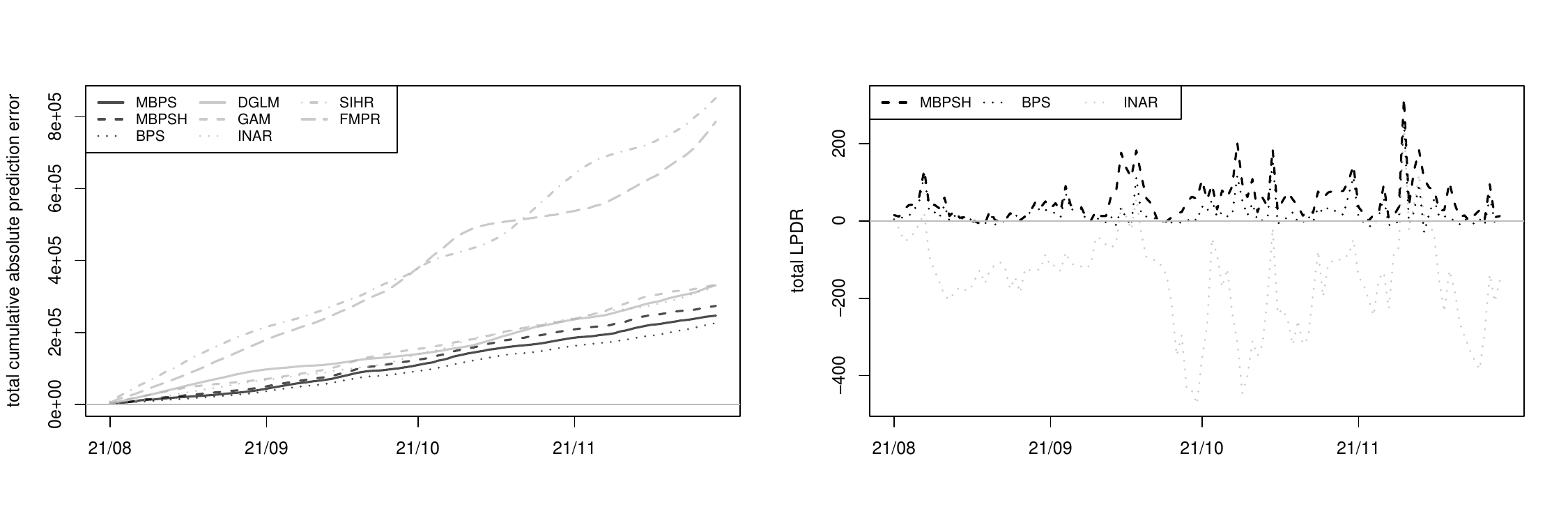}
\caption{One-, three- and seven-step-ahead total CAPE (left) and total LPDR (right) for the Korean data}
\label{fig:res_kr}
\end{figure}

Figure~\ref{fig:res_kr} shows the CAPE and LPDR summed over MAGs. 
For visibility, again, the total LPDR is shown only for MBPSH, BPS and INAR. 
In the cases of $s=1$ and $3$, the total CAPE for MBPS, MBPSH, BPS and INAR appear comparable.  
In the case of $s=7$, BPS resulted in a smaller total CAPE followed by MBPS and MBPSH. 
Similar to Figure~\ref{fig:res_seoul}, SIHR and FMPR are the models with the largest total CAPE. 
Turning to total LPDR, in the case of $s=1$, MBPSH and BPS are comparable throughout the prediction period. 
On the other hand, the total LPDR for INAR drops in the second half of the prediction period after October 2021. 
In the cases of $s=3$ and $7$, MBPSH tends to result in the highest total LPDR, followed by BPS. 
The total LPDR for INAR is negative for most of the prediction period, implying the superiority of the BPS models over INAR.

Table~\ref{tab:cov_kr} shows the coverages of 95\% prediction intervals.  
Most models resulted in undercoverage, especially in the case of $s=7$, again indicating the difficulty of accurately predicting isolated COVID-19 cases. 
Nonetheless, MBPSH resulted in the coverages closest to the target value of $0.95$ in all the cases, followed by BPS and MBPS. 
We also implemented the BPS models by inflating the variances of the latent factors by five, but we obtained similar results as in the case of the Japanese data. 
See also Appendix.

\begin{table}[H]
\caption{Coverages of 95\% prediction intervals for  one-, three-, seven-step-ahead prediction for the Korean data}
\label{tab:cov_kr}
\centering
\begin{tabular}{crrrrrrrr}\hline
  &MBPS & MBPSH &  BPS&  DGLM &  GAM & INAR & SIHR & FMPR\\\hline
$s=1$ & 0.918& 0.964& 0.932& 0.605& 0.169& 0.848& 0.028 &0.351\\
$s=3$ & 0.724& 0.944& 0.767& 0.590& 0.164& 0.654& 0.022 &0.354\\
$s=7$ & 0.637& 0.953& 0.697& 0.575& 0.169& 0.456& 0.028 &0.360\\
\hline
\end{tabular}
\end{table}

\subsubsection{Analysis of latent factors}
\begin{figure}[H]
\includegraphics[scale=0.37]{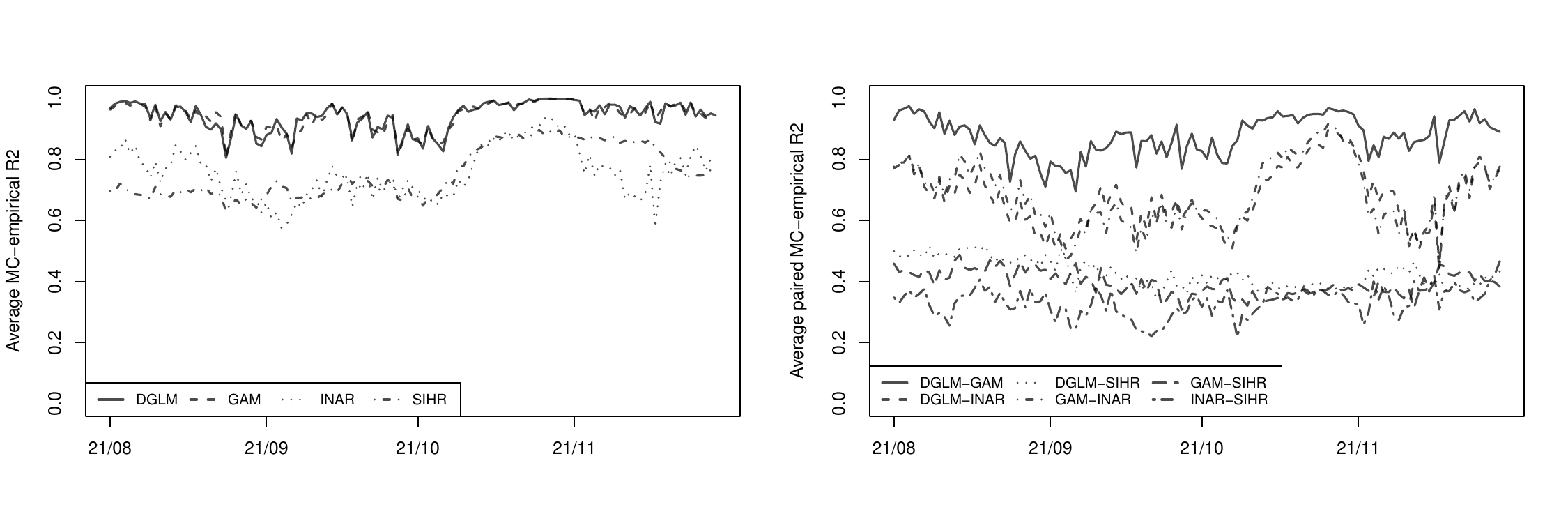}
\includegraphics[scale=0.37]{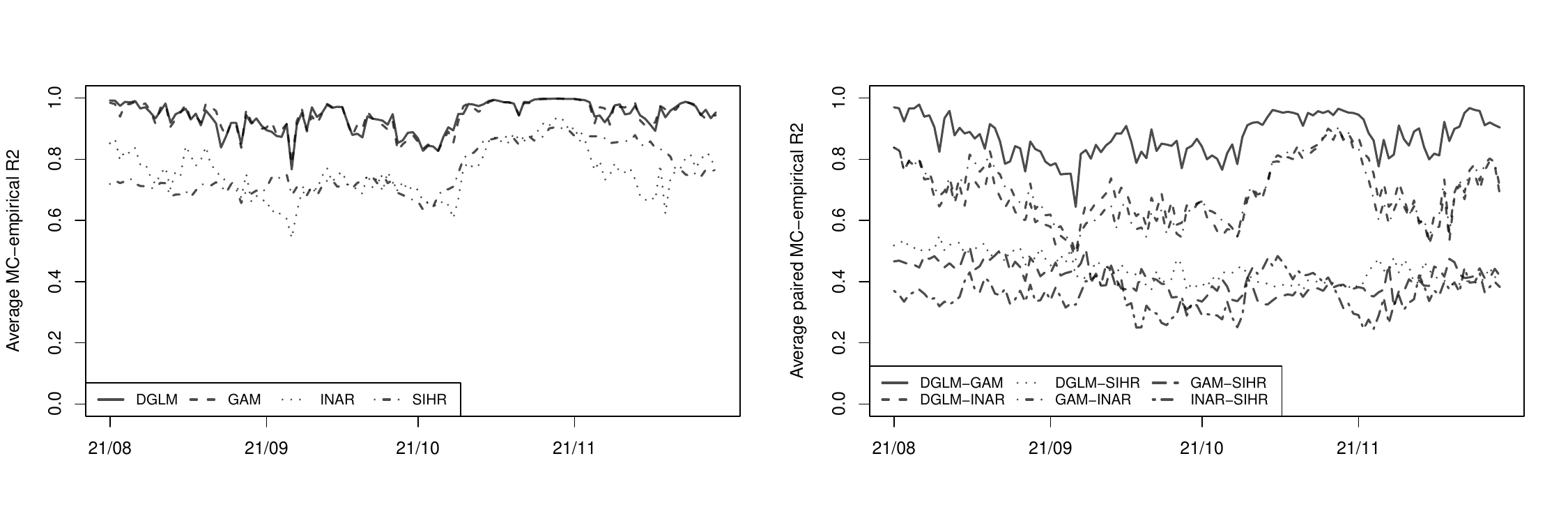}\\
\includegraphics[scale=0.37]{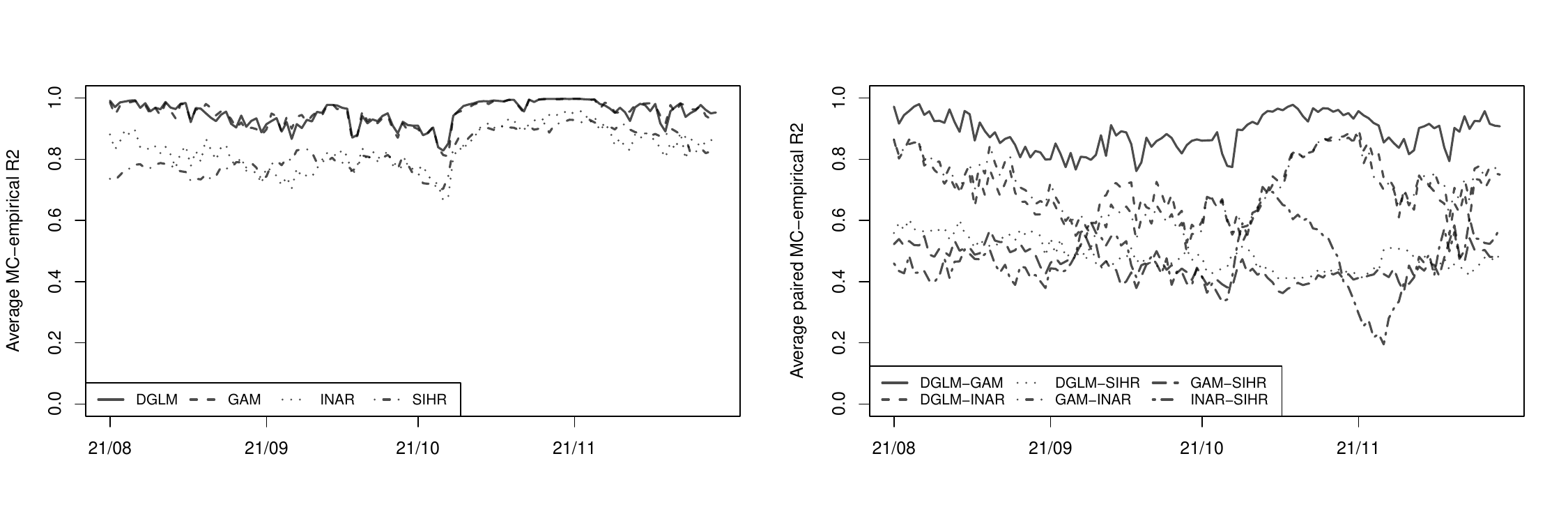}\\
\caption{MC-empirical $R^2$ (left) and paired MC-empirical $R^2$ (right) under MBPSH averaged over MAGs given all Korean data
}
\label{fig:res_total_kr_factors}
\end{figure}

Figure~\ref{fig:res_total_kr_factors} presents the MC-empirical $R^2$ and paired MC-empirical $R^2$ for one-, three- and seven-step ahead prediction averaged over MAGs given all Korean data up to up to  30 November 2021. 
The left panes show that the $R^2$ for DGLM and GAM are almost identical and are close to one, indicating high conditional dependence in all cases. 
The $R^2$ for INAR and SIHR are lower than those for DGLM and GAM. 
The levels of $R^2$ do not seem to vary over the different prediction horizons. 
It is also seen in the three- and seven-step ahead prediction.  
$R^2$ for all models dropped at the beginning of October 2021. 
In the right panels, we observe similar patterns in the paired $R^2$ over the different prediction horizons. 
The paired $R^2$ for the DGLM-GAM pair are also the highest of all pairs, as shown in the right panels. 
Those for the DGLM-INAR and GAM-INAR pairs exhibit similar behaviour, as GAM and DGLM produced similar prediction results shown in Figure~\ref{fig:res_kr}.

\begin{figure}[H]
\centering
\includegraphics[scale=0.33]{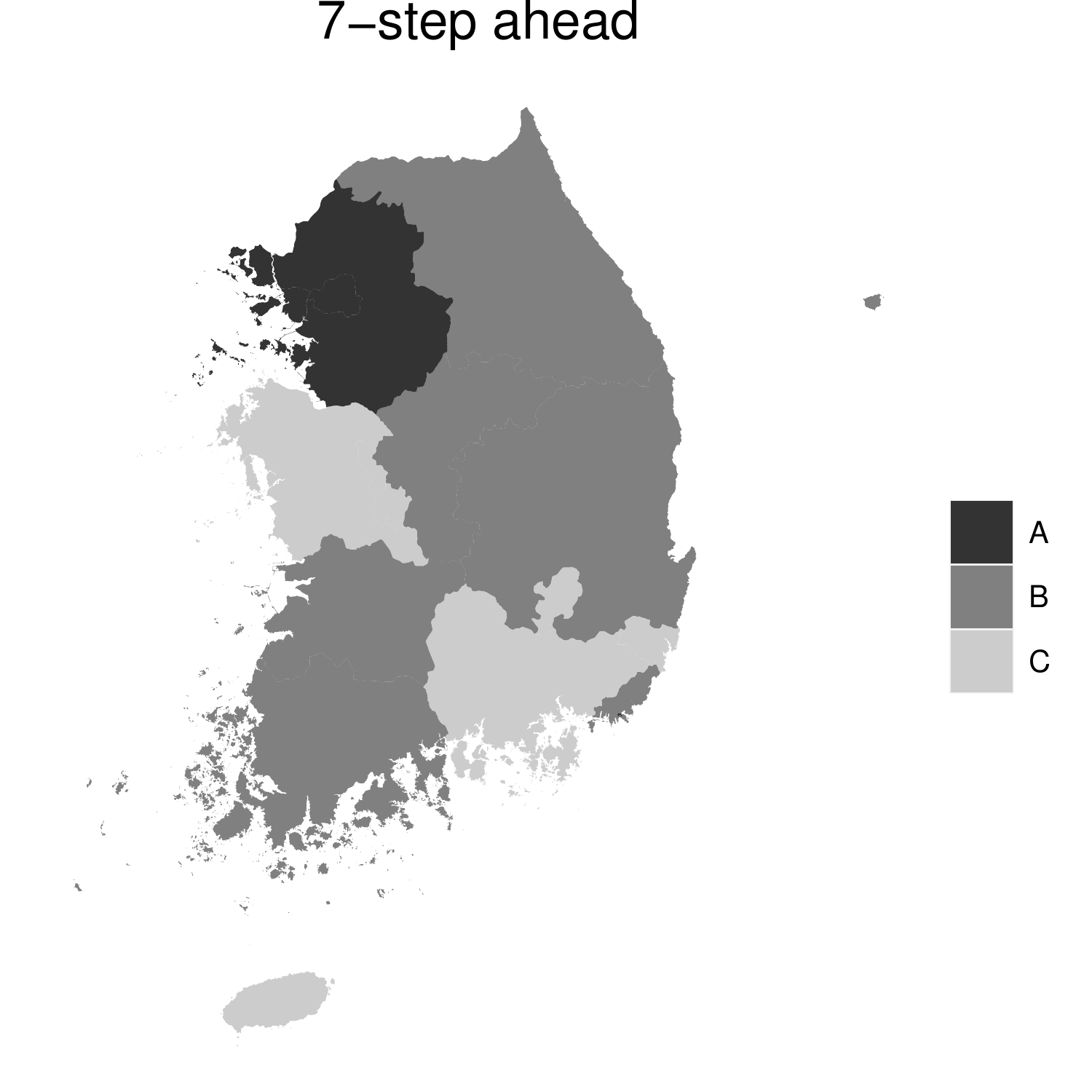}\\
\includegraphics[scale=0.55]{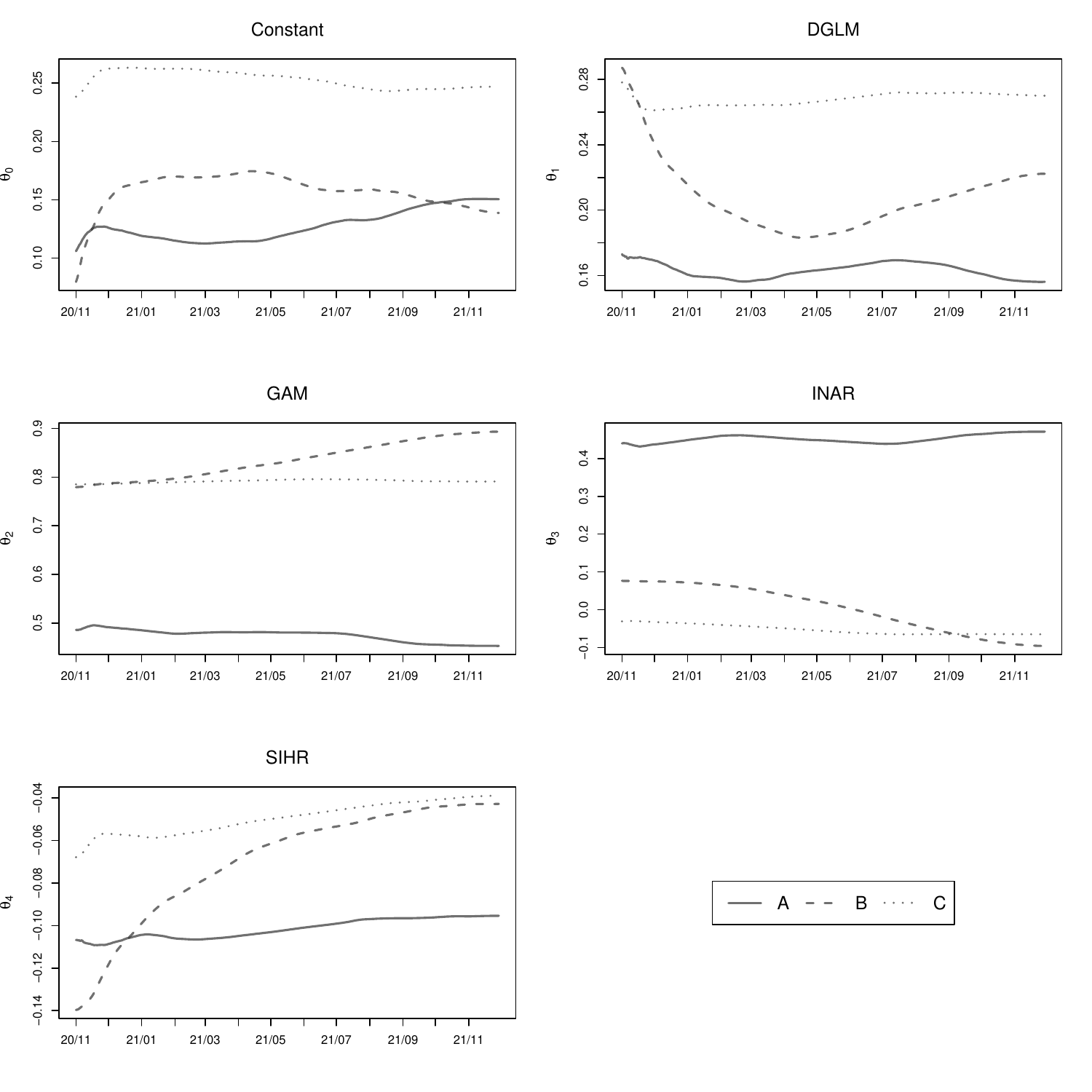}
\caption{Clustered MAGs (top maps)  and  posterior means of the synthesis weights under MBPSH for  seven-step ahead prediction (bottom panels) given all Korean data}
\label{fig:res_cluster_kr}
\end{figure}

\subsubsection{Clustering results}
Figure~\ref{fig:res_cluster_kr} presents the clustered MAGs and posterior means of the synthesis weights for $s=7$ based on the clustering results under MBPSH using all Korean data up to  30 November 2021. 
Due to space limitation, the results for $s=1$ and $3$ are presented in Appendix. 
It is shown that there are three clusters. 
Cluster~A includes Seoul and its neighbours, Incheon and Gyeonggi-do. 
The levels of the time series for these MAGs are the highest in our data, as shown in Figure~\ref{fig:kr}. 
Cluster~B includes Jeollabuk-do, Chungcheongbuk-do, Gangwon-do, Busan, Gyeongsangbuk-do and Jeollanam-do, whose levels are next to those of MAGs in Cluster~A. 
Therefore, the clusters are associated with the levels of the time series, as in the result of the Japanese data analysis in Section~\ref{sec:clust_jp}. 
We also observe a similar pattern in the case of $s=3$, while the pattern is unclear in the case of $s=1$.

In the bottom panels of Figure~\ref{fig:res_cluster_kr}, Cluster~C has the largest synthesis weight for DGLM. 
The trajectory of the weight for DGLM for Cluster~B is in a U-shape, with the weight decreasing towards May 2021 and increasing towards the end of the data period. 
For GAM, the synthesis weights are relatively high for all clusters, especially for Clusters~B and C. 
The weight for Cluster~B is increasing towards the end of the data period. 
On the contrary, for INAR, Cluster~A constantly has the highest synthesis weight, and the weight for Cluster~C is constantly close to zero. 
The synthesis weights for SIHR are relatively close to zero for all clusters. 

\begin{figure}[tb]
\centering
\includegraphics[scale=0.52]{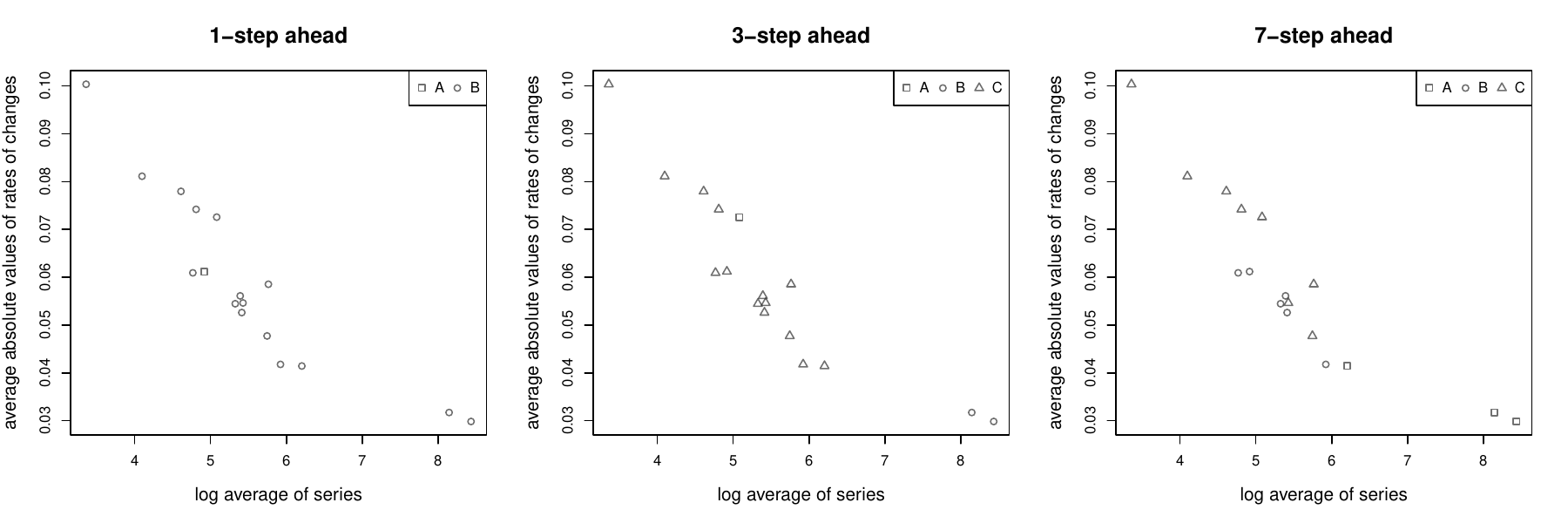}
\caption{Profile of the clustered time series given all  Korean data}
\label{fig:clusters_prof_kr}
\end{figure}

Figure~\ref{fig:clusters_prof_kr} presents the scatter plot of the log averages of the time series and average absolute values of the rates of changes from previous weeks in the time series. 
The locations of the series that belong to the same cluster are similar in the cases of $s=3$ and $7$. 
As in the observations above, the clusters are related to the levels of the time series. 
The figure also shows that time series with higher levels are associated with higher degrees of smoothness, such as the series in Cluster~A in the case of $s=7$ and those in Cluster~B in the case of $s=3$. 

\begin{figure}[H]
\centering
\includegraphics[scale=0.33]{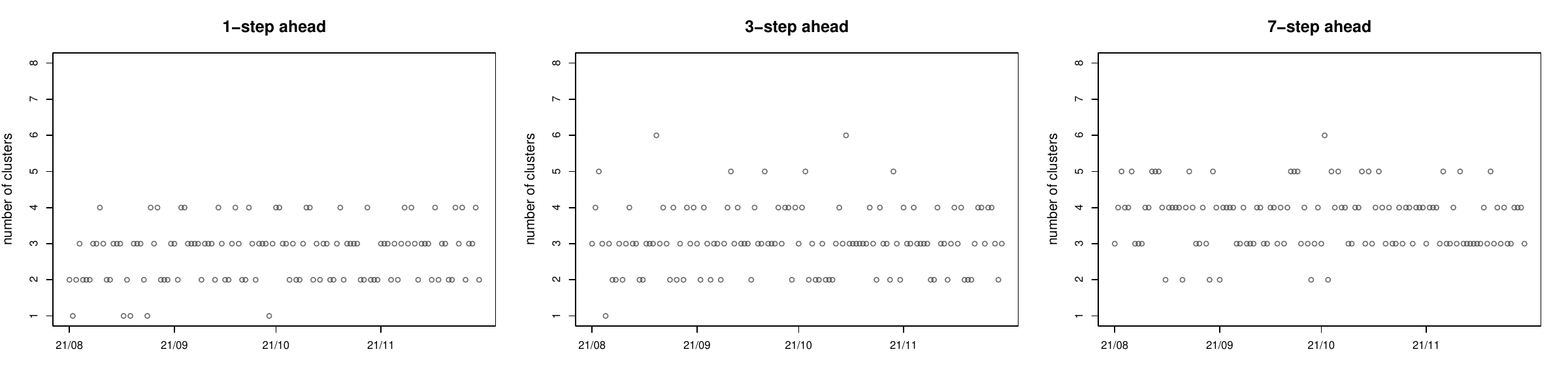}
\caption{Numbers of alive clusters under MBPSH for the Korean data}
\label{fig:num_clusters_kr}
\end{figure}

Finally, Figure~\ref{fig:num_clusters_kr} shows the numbers of clusters under MBPSH over the data period. 
As in the case of Japanese data, the numbers of clusters are stable over the data period. 
The average numbers of clusters are $2.7$, $3.1$ and $3.7$ for $s=1$, $3$ and $7$, respectively. 
Therefore, the number of clusters tends to increase with the prediction horizon.

\section{Conclusion}\label{sec:conc}
We have proposed the novel BPS methodology, the mixture of Bayesian predictive syntheses and variants, for multiple time series count data. 
The advantages of MBPS include the smaller number of BPS parameters and latent variables than the multivariate BPS and the use of univariate agent models. 
MBPS is particularly useful in multiple discrete time series analyses since the specification and implementation of a multivariate count model are generally cumbersome. 
The analysis of the Japanese and Korean data on the number of COVID-19 inpatients and isolated cases, respectively, showed that MBPS has improved predictive performance. 
The results also suggest that BPS  is an effective statistical methodology that provides vital information for policy-making against infectious diseases.

In the real data analysis, we experienced the sudden failure of an agent model while the synthesis weights remained high due to sudden surges and drops in the COVID-19  data.
This may amplify the deterioration of the predictive performance of MBPS. 
Since this problem is not specific to MBPS only, including a fail-safe mechanism within BPS would improve the performance of the BPS methodology and expand its applicability. 
We would like to explore the possible extension of BPS in this direction in future studies.

\bibliographystyle{chicago}
\bibliography{Ref}

\appendix

\section{Posterior computation}\label{sec:mcmc}
\subsection{Gibbs sampler for MBPS}
The posterior inference is carried out by using the Markov chain Monte Carlo (MCMC) method. 
In order to facilitate the MCMC sampling, we follow the approach of \cite{HIS21} and \cite{Dangelo} to approximate the Poisson pmf by the negative binomial pmf $\tilde{\alpha}(y_{it}|\f_{it},\bthe_{tk},r)$ with a large dispersion parameter $r$ and then apply P\'olya-gamma (PG) mixture of \cite{POLSON13}. 
For a sufficiently large $r$, we have
\begin{eqnarray*}
\alpha(y_{it}|\f_{it},\bthe_{tk})&\approx&\tilde{\alpha}(y_{it}|\f_{it},\bthe_{tk},r)\\
&=&\frac{\Gamma(y_{it}+r)}{\Gamma(r)y_{it}!}\frac{(e^{\psi_{itk}})^{y_{it}}}{(1+e^{\psi_{itk}})^{y_{it}+r}}\\
&=&\frac{\Gamma(y_{it}+r)}{\Gamma(r)y_{it}!}2^{-b_{it}}\exp\left\{\kappa_{it}\psi_{itk}\right\}\int_0^\infty \exp\left\{-\frac{\omega_{itk}\psi_{itk}^2}{2}\right\}p(\omega_{itk}|b_{it},0)d \omega_{itk},
\end{eqnarray*}
where $b_{it}=y_{it}+r$, $\kappa_{it}=(y_{it}-r)/2$, $\psi_{itk}=\bthe_{tk}'\F_{it}-\log r$ and $\omega_{itk}$ follows the PG distribution $PG(b_{it},0)$ with the density $p(\omega_{itk}|b_{it},0)$. 
Throughout this paper, $r=1000$ is used. 
See also \cite{Dangelo} for a data-driven approach to determine the value of $r$. 

Then the joint distribution of the parameters and latent variables for $t=1,\dots,T$ is proportional to
\begin{equation}\label{eqn:joint}
\begin{split}
& p(\bpi)
\prod_{t=1}^Tp(\bthe_{tk}|\bthe_{t-1,k},\bSig_{tk})\\
&\qquad\times\left(\prod_{i=1}^n\left[\pi_k\exp\left\{\kappa_{it}\psi_{itk}\right\} \exp\left\{-\frac{\omega_{itk}\psi_{itk}^2}{2}\right\}p(\omega_{itk}|b_{it},0)\right]^{I(z_i=k)}
\prod_{j=1}^Jh_{itj}(f_{itj})\right),
\end{split}
\end{equation}
where $p(\bpi)$ is the prior density for $\bpi$ and $I(\cdot)$ is the indicator function.  
Then, our Gibbs sampler alternately samples $\left\{z_i\right\}$,  $\left\{\bthe_{tk}\right\}$, $\{{\f}_{it}\}$,  $\left\{\omega_{itk}\right\}$, $\left\{s_{ik}\right\}$ and $\bpi$. 

\begin{description}
\item[Sampling $z_{i}$:]
For $i=1,\dots,n$, $z_i$ is sampled from the categorical distribution with probabilities proportional to
$\Pr(z_i=k|\text{Rest})\propto\pi_k\prod_{t=1}^T{\alpha}(y_{it}|\f_{it},\bthe_{tk},r)$ for $k=1,\dots,K$.
\item[Sampling $\bpi$:]
The full conditional distribution of $\bpi$ is the Dirichlet distribution  $Dir(a_0+\sum_is_{i1},\dots,a_0+\sum_is_{iK})$.

\item[Sampling $\omega_{itk}$:]
For $i=1,\dots,n$, $t=1,\dots,T$, $\omega_{itz_i}$ is sampled from $PG(r+y_{it},\bthe_{tz_i}'\F_{it}-\log r)$.

\item[Sampling ${\f}_{it}$:]
Let us define  ${\bthe}_{-0,tk}=(\theta_{tk1},\dots,\theta_{tkJ})$. 
From \eqref{eqn:joint}, the full conditional distribution of ${\f}_{it}$ is proportional to
\[
\begin{split}
p({\f}_{it}|\text{Rest})&\propto \exp\left\{\kappa_{it}\psi_{itz_i}\right\} \exp\left\{-\frac{\omega_{itz_i}\psi_{itz_i}^2}{2}\right\}\prod_{j=1}^Jh_{itj}(f_{itj})\nonumber\\
&\propto\exp\left\{-\frac{1}{2}\left({\f}_{it}'\omega_{itz_i}{\bthe}_{-0,tz_i}{\bthe}_{-0,tz_i}'{\f}_{it} - 2(\omega_{itz_i}(\log r-\theta_{tz_i0}) + \kappa_{it}){\f}_{it}'{\bthe}_{-0,tz_i})\right)\right\}\\
&\quad\times \prod_{j=1}^Jh_{itj}(f_{itj})
\end{split}
\]
Then, assuming that $h_{itj}(\cdot)$ is given by the density of $N(m_{itj},s^2_{itj})$ for $j=1,\dots,J$, ${\f}_{it}$ is sampled from $N(\hat{\m}_{it},\hat{\S}_{it})$ where
\[
\hat{\m}_{it}=\hat{\S}_{it}\left[\left(\omega_{itz_i}(\log r-\theta_{tz_i0}) + \frac{y_{it}-r}{2}\right){\bthe}_{-0,tz_i} + \S_{it}^{-1}\m_{it}\right],\quad
\hat{\S}_{it}= \left[\omega_{itz_i}{\bthe}_{-0,tz_i}{\bthe}_{-0,tz_i}'+\S_{it}^{-1}\right]^{-1}, 
\]
for $i=1,\dots,n$, $t=1,\dots,T$, where  $\m_{it}=(m_{it1},\dots,m_{itJ})'$ and $\S_{it}=\diag(s_{it1}^2,\dots,s_{itJ}^2)$.

\item[Sampling $\bthe_{tk}$:]
Let us define $n_k=\sum_{i=1}^nI(z_i=k)$, which is the number of time series belonging to $k$th cluster. 
From \eqref{eqn:joint}, we have 
\[
p(\bthe_{1:T,k}|\text{Rest})\propto\prod_{t=1}^T
\exp\left\{-\frac{1}{2}(\d_{tk}-\F_{tk}\bthe_{tk})'\bOmega_{tk}(\d_{tk}-\F_{tk}\bthe_{tk})\right\}
p(\bthe_{tk}|\bthe_{t-1,k},\bSig_{tk}),
\]
where for $i$ such that $z_i=k$, the $n_k\times 1$ vector $\d_{tk}$ is the collection of $(y_{it}-r)/2\omega_{itk}+\log r$, the rows of the $n_k\times (J+1)$ matrix $\F_{tk}$ consists of $\F_{it}'$'s and the $n_k\times n_k$ diagonal matrix $\bOmega_{tk}$ has $\omega_{itk}$'s on the diagonal. 
This can be recognised as the joint distribution of the state vectors $\bthe_{tk}$ and observed data $\d_{tk}$ for the Gaussian linear state space model given by
\[
\begin{split}
\d_{tk}&=\F_{tk}\bthe_{tk} + \v_{tk},\quad \v_{tk}\sim N(\zero,\bOmega_{tk}^{-1}),\\
\bthe_{tk}&=\bthe_{t-1,k} + \e_{tk},\quad \e_{tk}\sim N(\zero,\bSig_{tk}), 
\end{split}
\]
with the discount factor $\delta_\bSig$ for $\bSig_{tk}$. 
Then given the values for $\d_{tk}$, $\F_{tk}$, $\bOmega_{tk}$ and $\bthe_{1:T,k}$ can be sampled sequentially using the forward filtering and backward sampling (FFBS) described in the following.
We introduce the common discount factor $\delta_\bSig\in(0,1]$ for $\bSig_{tk}$. 

\begin{itemize}
    \item
    Forward filtering:
    \begin{enumerate}
        \item 
        At the time $t-1$, the posterior distribution of $\bthe_{t-1,k}$ is given by
        \[
        \bthe_{tk}|\d_{1:t-1,k},\F_{1:t-1,k}\sim N(\m_{t-1,k},\C_{t-1,k}).
        \]
        \item
        The prior distribution of $\bthe_{tk}$ at the time $t$ is 
        $
            \bthe_{tk}|\d_{1:t-1,k},\F_{1:t-1,k}\sim N(\a_{tk},\R_{tk}),
        $
        where $\a_{tk}=\m_{t-1,k}$, $\R_{tk}=\C_{t-1,k}/\delta_\bSig$. 
        \item
        The one-step ahead predictive distribution of $\d_{tk}$ is given by
        $
            \d_{tk}|\d_{1:t-1,k},\F_{1:t,k}\sim N(\g_t,\Q_t)
        $
        where $\g_t=\F_{tk}'\a_{tk}$ and $\Q_{tk}=\F_{tk}'\R_{tk}\F_{tk}+\bOmega^{-1}_{tk}$. 
        \item
        The posterior distribution of $\bthe_{tk}$ at the time $t$ is given by $\bthe_{tk}|\d_{1:t,k},\F_{1:t,k}\sim N(\m_{tk},\C_{tk})$ where 
        $\m_{tk}=\a_{tk} + \A_{tk}(\d_{tk}-\g_{tk})$, $\C_{tk}=\R_{tk}-\A_{tk}\Q_{tk}\A_{tk}'$ and $\A_{tk}=\R_{tk}\F_{tk}\Q_{tk}^{-1}$. 
        
    \end{enumerate}
    \item
    Backward sampling:
    \begin{enumerate}
        \item 
        At the time $T$,  sample from $\bthe_{Tk}|\d_{1:T,k},\F_{1:T,k}\sim  N(\m_{Tk},\C_{Tk})$. 
        \item
        For $t=T-1,T-2,\dots,1$, $\bthe_{tk}$ is sampled from the conditional distribution $\bthe_{tk}|\bthe_{t+1,k},\d_{1:T,k},\F_{1:T,k}\sim N(\m^*_{tk},\C^*_{tk})$ where $\m_{tk}^*=\m_{tk}+\delta_\bSig(\bthe_{t+1,k}-\a_{t+1,k})$ and $\C^*_{tk}=(1-\delta_\bSig)\C_{tk}$. 
    \end{enumerate}
\end{itemize}


\end{description}

\subsection{Sampling steps for MBPSH}
The sampling steps for MBPSH is almost the same as those for MBPS, but with some small modification due to the added term $u_{it}$. 
\begin{description}
    \item[Sampling $\bthe_{tk}$ and $\tilde{f}_{it}$:]
    The same sampling methods described above can be used, but now with the vector $\d_{tk}$ being the collection of $ (y_{it}-r)/2\omega_{itk}+\log r - u_{it}$. 
    \item[Sampling $u_{it}$:]
    For $i=1,\dots,N$ and $t=1,\dots,T$, the full conditional distribution of  $u_{it}$ is $N(\hat{u}_{it},\hat{v}^2_{it})$ where
    \[
        \hat{u}_{it}=\hat{v}^2_{it}\left(\omega_{itz_i}(\log r - \bthe_{tz_i}'\F_{it})+ \frac{y_{it} - r}{2}\right),\quad
        \hat{v}^2_{it}=\left(\omega_{itz_i}+\frac{1}{\tau_{tk}^2}\right)^{-1}.
    \]
    \item [Sampling $\tau_{tk}^2$:]
    The FFBS algorithm is used for the precision parameter $\varphi^2_{tk}=1/\tau_{tk}^2$ for $k=1,\dots,K$. 
    \begin{itemize}
        \item Forward filtering:
        \begin{enumerate}
            \item 
            At the time $t-1$, the posterior distribution of $\varphi_{t-1,k}^2$ is given by the gamma distribution $Ga(a_{t-1}/2,b_{t-1}/2)$. 
            \item
            The prior distribution a the time $t$ is $Ga(\beta_\tau a_{t-1}/2,\beta_\tau b_{t-1}/2)$. 
            \item 
            At the time $t$, the posterior distribution is given by $Ga(a_{t}/2,b_{t}/2)$, where $a_t=\beta_\tau a_{t-1}+n_k$ and $b_t=\beta_\tau b_{t-1} + \sum_{i:z_i=k}u_{it}^2$. 
            
        \end{enumerate}
        \item Backward sampling: If $t=T$, draw $\varphi_{tk}^2$ from $Ga(a_T/2,b_T/2)$, otherwise, $\phi_{tk}^2=\varphi_{t+1,k}^2+e_{tk}$ where $e_{tk}\sim Ga((1-\beta_\tau)a_t/2,b_t/2)$, for $t=T-1,\dots,1$.

    \end{itemize}

\end{description}

\subsection{SIHR compartment model}
The SIHR compartment model is given by the following system of differential equations:
\begin{equation*}
\begin{split}
    dS/dt &= -\alpha I \cdot S / N\\
    dI/dt &= \alpha I\cdot S / N- (\beta + \delta_I)  I\\
    dH/dt &= \beta  I - \delta_H  H\\
    dR/dt &= \delta_I I  + \delta_H H
\end{split}
\end{equation*}
where $S$, $I$, $H$ and $R$ are the numbers of susceptible, infected, hospitalised and recovered individuals, respectively, $\alpha, \beta, \delta_I, \delta_H\in(0,1)$ are the parameters representing the infection rate, hospitalisation rate, recovering rate from the infected and hospitalised states, respectively. 
The Poisson means are modelled as the solution of the differential equations for $H$.

\subsection{Coverages with the inflated variances}

\begin{table}[H]
\caption{Coverages of 95\% prediction intervals for one-step-ahead prediction for Japanese and Korean data under the BPS models with the variances of the latent factors increased by five times}
\label{tab:cov_inf}
\centering
\begin{tabular}{crrrr}\toprule
Country &  $s$ &   MBPS &  MBPSH & BPS \\\hline
Japan &1 & 0.559 &0.929 & 0.582\\
Korea &1 & 0.910 & 0.969& 0.936\\
 &3 & 0.725 & 0.946 &0.770\\
 &7 & 0.639 &0.949& 0.696\\
\bottomrule
\end{tabular}
\end{table}

\subsection{Clustering results for Korean data}
\begin{figure}[H]
\centering
\includegraphics[scale=0.3]{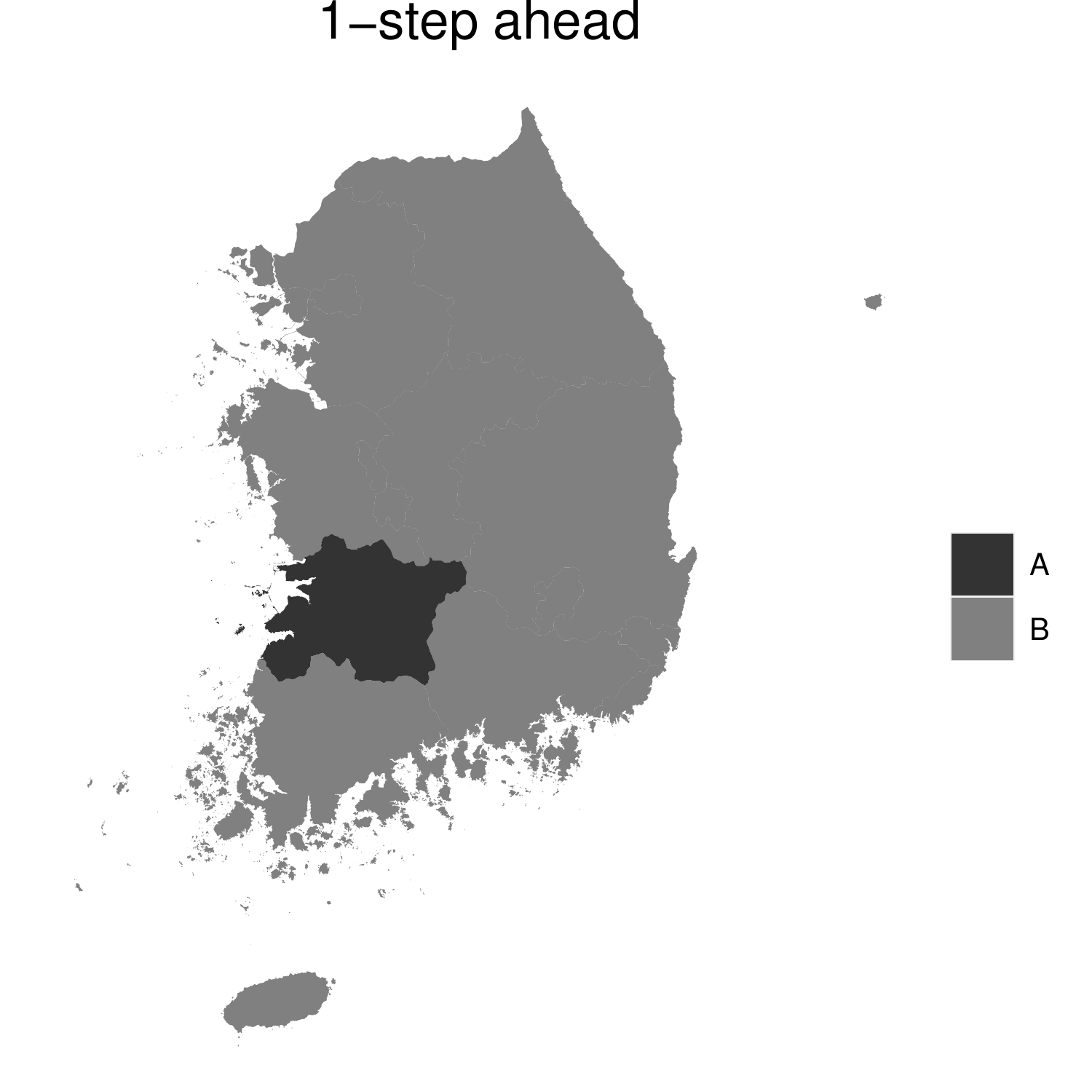}\\
\includegraphics[width=\textwidth]{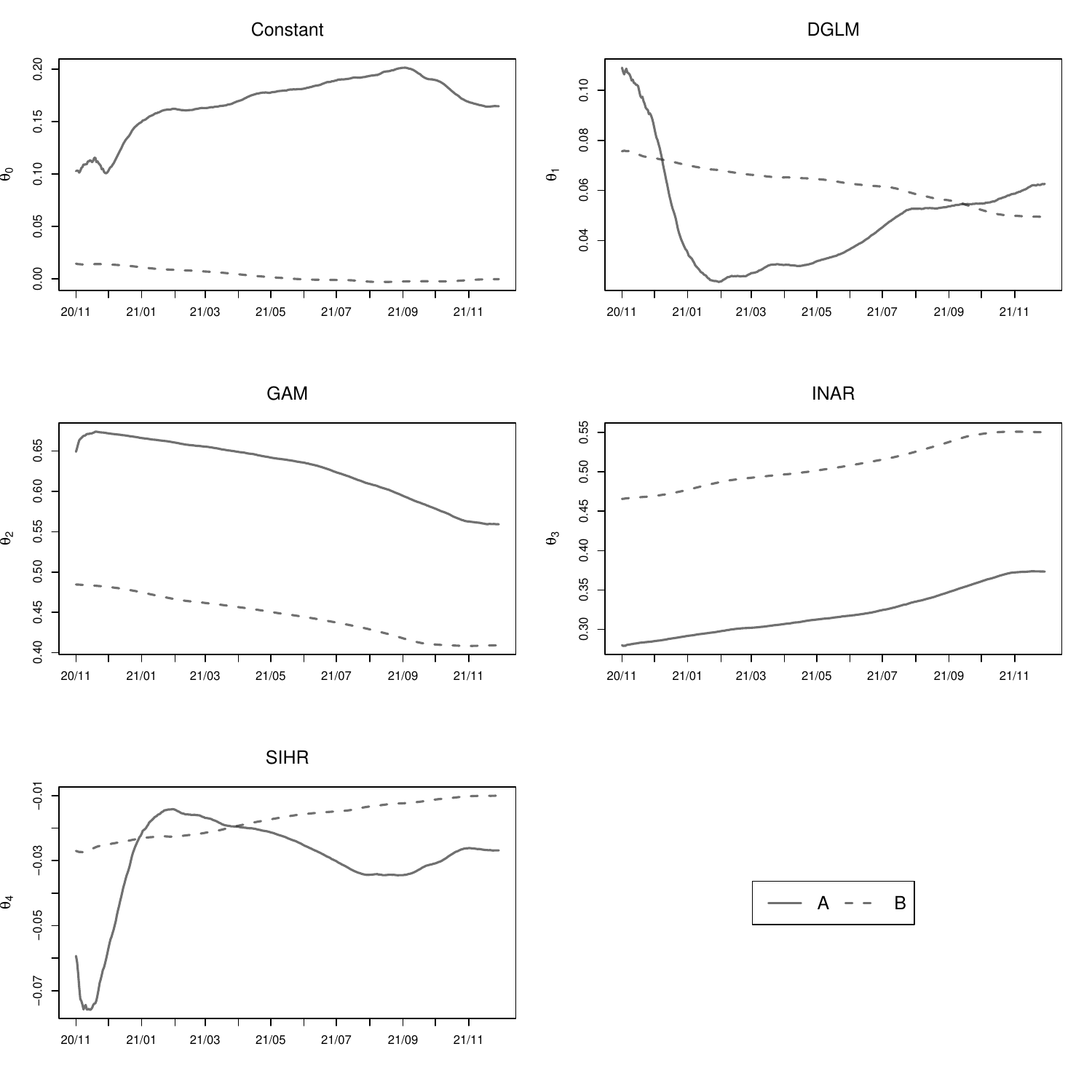}
\caption{Clustered MAG (top maps)  and  posterior means of the synthesis weights  under MBPSH for  one-step ahead prediction given all Korean data}
\label{fig:res_cluster_kr1}
\end{figure}

\begin{figure}[H]
\centering
\includegraphics[scale=0.3]{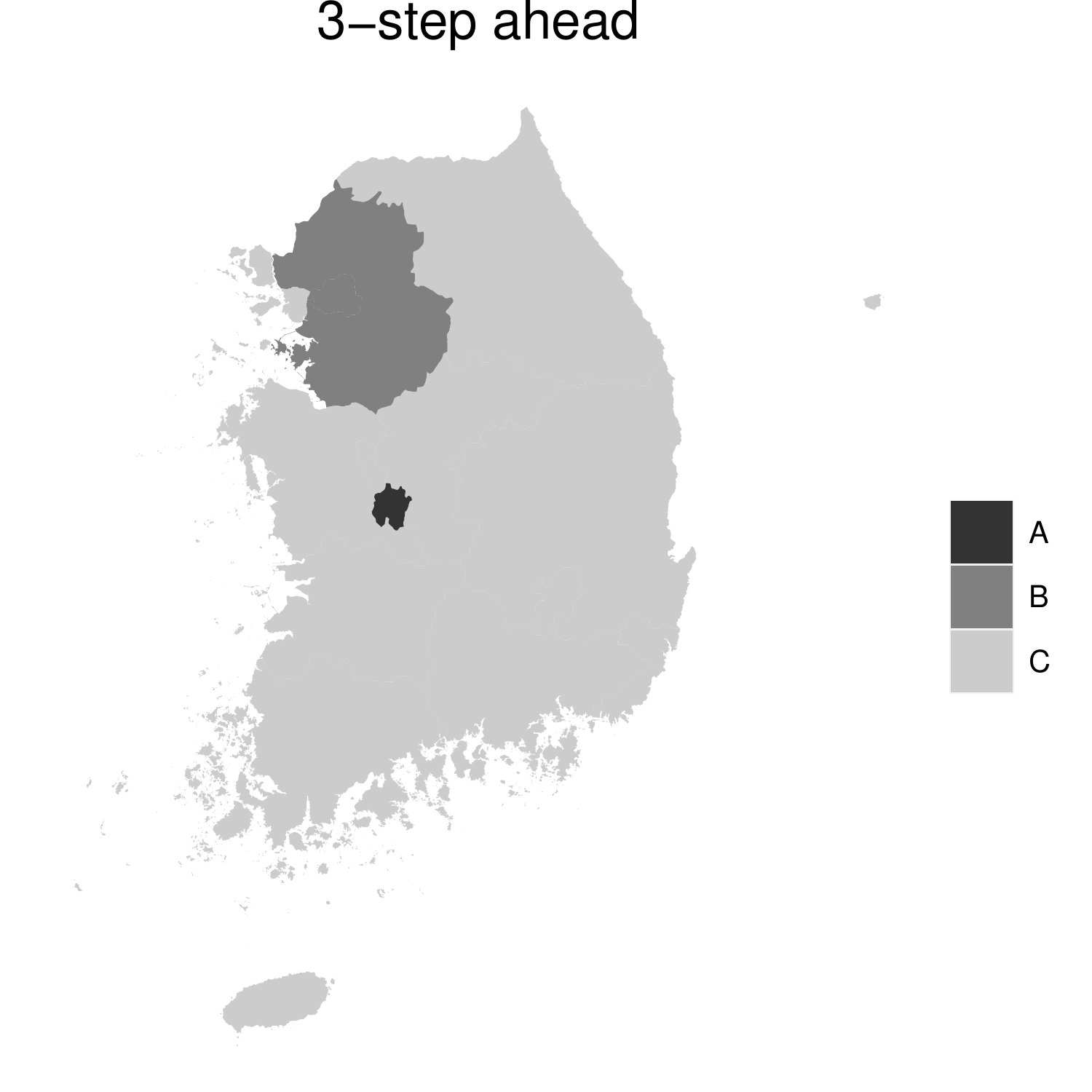}\\
\includegraphics[width=\textwidth]{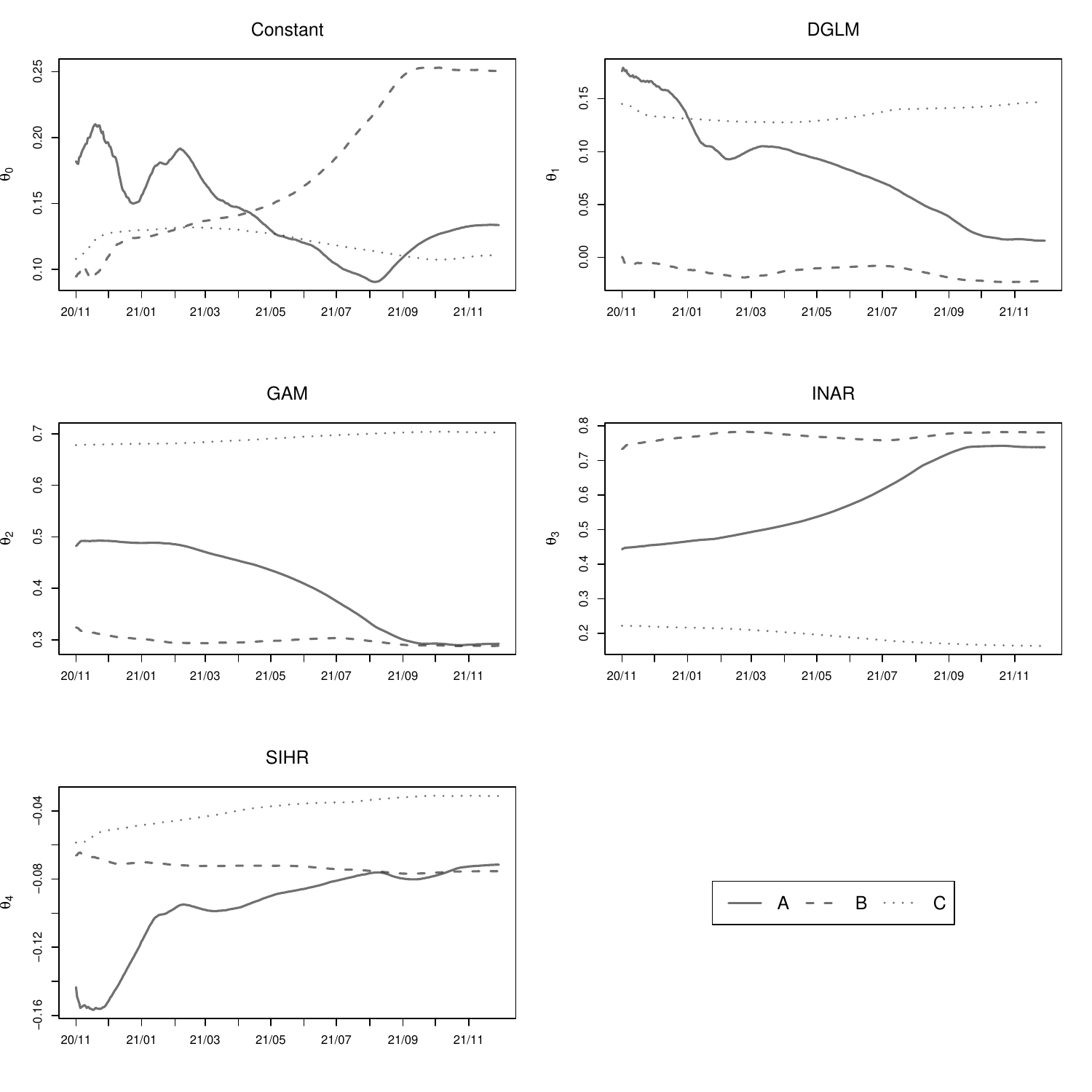}
\caption{Clustered MAG (top maps)  and  posterior means of the synthesis weights  under MBPSH for  three-step ahead prediction given all Korean data}
\label{fig:res_cluster_kr2}
\end{figure}

\end{document}